\documentclass[12pt]{article}
\usepackage{amsmath}
\usepackage[margin=2.0cm]{geometry}
\usepackage{epsfig}
\usepackage{cite}
\usepackage{amsfonts}
\usepackage{fancyhdr}
\usepackage{setspace}
\usepackage{enumitem}
\usepackage{graphicx}
\usepackage{subcaption}
\usepackage[font={sf,sl}]{caption}
\usepackage{bm}
\usepackage{cancel}
\usepackage{wasysym}
\usepackage{amssymb}
\usepackage{wrapfig}
\usepackage{xcolor}
\usepackage{textcomp}
\usepackage{arydshln}
\usepackage{tikz}
\usepackage{authblk}
\usepackage{titlesec}
\usepackage{hyperref}
\usepackage{textcmds}
\usepackage[export]{adjustbox}
\usepackage[page,toc,titletoc,title]{appendix}
\usetikzlibrary{arrows}
\newcommand{\authorname}{Vineet Dawara} 
\newcommand{\Tp}{\ensuremath{T_{+}}}
\newcommand{\Tm}{\ensuremath{T_{-}}}
\newcommand{\To}{\ensuremath{T_{0}}}

\titleformat{\section}
{\normalfont\bfseries\scshape}{\thesection}{1em}{}

\titleformat{\subsection}
{\normalfont\bfseries\scshape}{\thesubsection}{1em}{}

\titleformat{\subsubsection}
{\normalfont\slshape}{\thesubsubsection}{1em}{}

\titleformat{\title}{\normalfont\bfseries}{\thesection}{1em}{}

\title{\vspace{-3em} \large \bfseries On the onset of slip at adhesive elastic interfaces}
\author[1]{\normalsize \authorname}
\author[1]{\normalsize Koushik Viswanathan\thanks{koushik@iisc.ac.in}}
\affil[1]{Department of Mechanical Engineering, Indian Institute of Science, Bangalore, India}

\date{\vspace{-1em}\normalsize \textsc{\today}}
\numberwithin{equation}{section}
\numberwithin{table}{section}
\begin{document}

\maketitle
\thispagestyle{plain}
\vspace{-1em}
\noindent\hrulefill
\begin{abstract}
  %

  
The transition from static to dynamic friction when an elastic body is slid over another is now known to result from the motion of interface rupture fronts. These fronts may be either crack-like or pulse-like, with the latter involving reattachment in the wake of the front. How and why these fronts occur remains a subject of active theoretical and experimental investigation, given its wide ranging implications for a range of problems in tribology. In this work, we investigate this question using an elastic lattice-network representation; bulk and interface bonds are simulated to deform and, in the latter case, break and reform dynamically in response to an applied remote displacement. We find that, contrary to the oft-cited rigid body scenario with Coulomb-type friction laws, the type of rupture front observed depends intimately on the location of the applied boundary condition. Depending on whether the sliding solid is pulled, pushed or sheared---all equivalent applications in the rigid case---distinct interface rupture modes can occur. We quantify these rupture modes, evaluate the interface stresses that lead to their formation, and and study their subsequent propagation dynamics. A strong analogy between the sliding friction problem and mode II fracture emerges from our results, with attendant wave speeds ranging from slow to Rayleigh. We discuss how these fronts mediate interface motion and implications for the general transition mechanism from static to dynamic friction.
\end{abstract}
\paragraph*{Keywords -} spring lattice model, friction, pulse modes, rupture fronts

\clearpage
\section{Introduction}
The question of how a stationary body begins to slide under an applied load is simple enough to state, yet remains the subject of significant current investigation. Understanding the origin and nature of the slip process has wide-ranging practical consqeuences, from sliding mechanical elements \cite{rabinowicz1956stick, rabinowicz1958intrinsic} to earthquake dynamics \cite{brace1966stick, scholz1998earthquakes}. If the sliding body is assumed to be perfectly rigid, then the transition from stationary to sliding is instantaneous---the well-known Amontons-Coulomb friction model predicts that this change occurs when the applied load exceeds the static friction threshold \cite{bowden2001friction}. While more complex friction models can help describe phenomena such as stick-slip \cite{dieterich1979modeling,  dieterich1978time,  ruina1983slip}, the underlying rigid body assumption implies that they too predict an instantaneous stationary to slip transition beyond a threshold.

To comprehend the limitations of this idealization and the actual temporal nature of the slip process, it is important to first identify the critical time-scale in the problem. For a solid being slid at speed $v_0$ and with interface/contact length $L$ in the sliding direction, the time taken for a disturbance to traverse the entire interface, termed the slip time $t_s$, is given by $t_s = L/v_s$. Here $v_s$ represents a typical elastic interface wave speed, such as the generalized Rayleigh wave speed \cite{achenbach1967dynamic}. If $t_s$ is very small compared to the experimental observation time, traversal of an interface disturbance appears instantaneous; $t_s$ is clearly increased either when $L$ is large, as in earthquake faults that span several 100 kilometers, or when $v_s$ is small, as happens when the sliding body can no longer be assumed rigid. Under these conditions, the transition from stationary to slip occurs via the interface's elastic compliance and must be considered carefully.

Over the past few decades, several high-speed imaging experiments of sliding interfaces have revealed rich underlying dynamics at the onset of slip. Local interface rupture fronts of several types have now been identified to be responsible for mediating slip \cite{rubinstein2004detachment,gvirtzman2021nucleation,xia2004laboratory,lu2007pulse, viswanathan2016stick}. Macroscopic sliding begins when these fronts traverse the entire interface length. Surprisingly, however, rupture (alternatively slip or detachment) fronts have been reported to propagate with a wide range of velocities---ranging from very slow \cite{ben2010dynamics, viswanathan2016stick} to sub-Rayleigh \cite{rubinstein2004detachment} to super-shear \cite{xia2004laboratory}. Additionally, such rupture fronts may broadly be delineated into two distinct classes---crack-like and pulse-like \cite{lu2007pulse, shlomai2016structure}. In the former, a single interface rupture front traverses the entire interface, following which (often steady) sliding ensues. In stark contrast, pulse-like modes exhibit readhesion behind the leading interface disturbance, causing seizure of motion in their wake \cite{mindlin1949compliance, johnson1971surface, savkoor1977effect, adams2014stick}. Which one of these modes occurs when transitioning from static to dynamic sliding remains an open question; hypotheses range from the need for velocity-weaking friction \cite{heaton1990evidence, zheng1998conditions}, stress barriers \cite{johnson1990initiation} or normal stress variations along the interface \cite{andrews1997wrinkle, cochard2000fault}. 

In order to address this question more comprehensively, we must reconsider the nature of elastic compliance. When a solid is being slid, say from left to right, the applied force is towards the right, with the opposing friction force acting to the left. If the solid is assumed perfectly rigid, the position of the applied load---whether it is shear on the top face, a push load on the left or a pull on the right---does not matter. However, for an elastic solid, the pre-sliding stress state corresponding to these different load applications will vary fundamentally, each representing a unique boundary value problem. Naturally, the nature of slip onset is also expected to differ significantly. Yet, there have been surprisingly few systematic investigations of this effect~\cite{tromborg2011transition, thogersen2021minimal}. Depending on the nature of the load and the contact stress distribution, rupture fronts may nucleate at either end of the contact, move in different directions, and potentially at different speeds \cite{viswanathan2016stick, gvirtzman2021nucleation}. This behaviour simply cannot be captured by 1D (or quasi 1D) models of the interface \cite{gerde2001friction, braun2009dynamics} 

In the present work, we systematically explore the role of boundary conditions in slip initiation and interface sliding using a numerical lattice model of the elastic bodies. This study builds on prior experimental investigations and analytical calculations that have been published elsewhere \cite{viswanathan2016stick, ansari2022propagating}. The model may be placed in the same category as the classical Burridge-Knopoff model and its relatives \cite{burridge1967model, amundsen20121d, braun2009dynamics} but differs in that it accounts for displacement perpendicular to the interface. This is a necessary (but not sufficient) condition for detachment waves to occur \cite{schallamach1971does, viswanathan2022fifty}. Our investigations reveal two things---the occurrence of three different rupture modes depending on the boundary conditions, and intimate connections between these and interface cracks that have been experimentally postulated before \cite{svetlizky2017brittle}

The manuscript is organized as follows. Details of our model and corresponding numerical methods are described in Sec.~\ref{sec:model}. Our primary findings, organized by type of loading, are presented in Sec.~\ref{sec:results}, and contain, especially, a detailed analysis of resulting interface slip dynamics (Sec.~\ref{sec:slip}). We place our results in a broader context and discuss some of their consequences in Sec.~\ref{sec:discussion}.

\section{Methods}\label{sec:model}

\begin{figure}[ht!]
	\centering	
	\includegraphics[width=\linewidth]{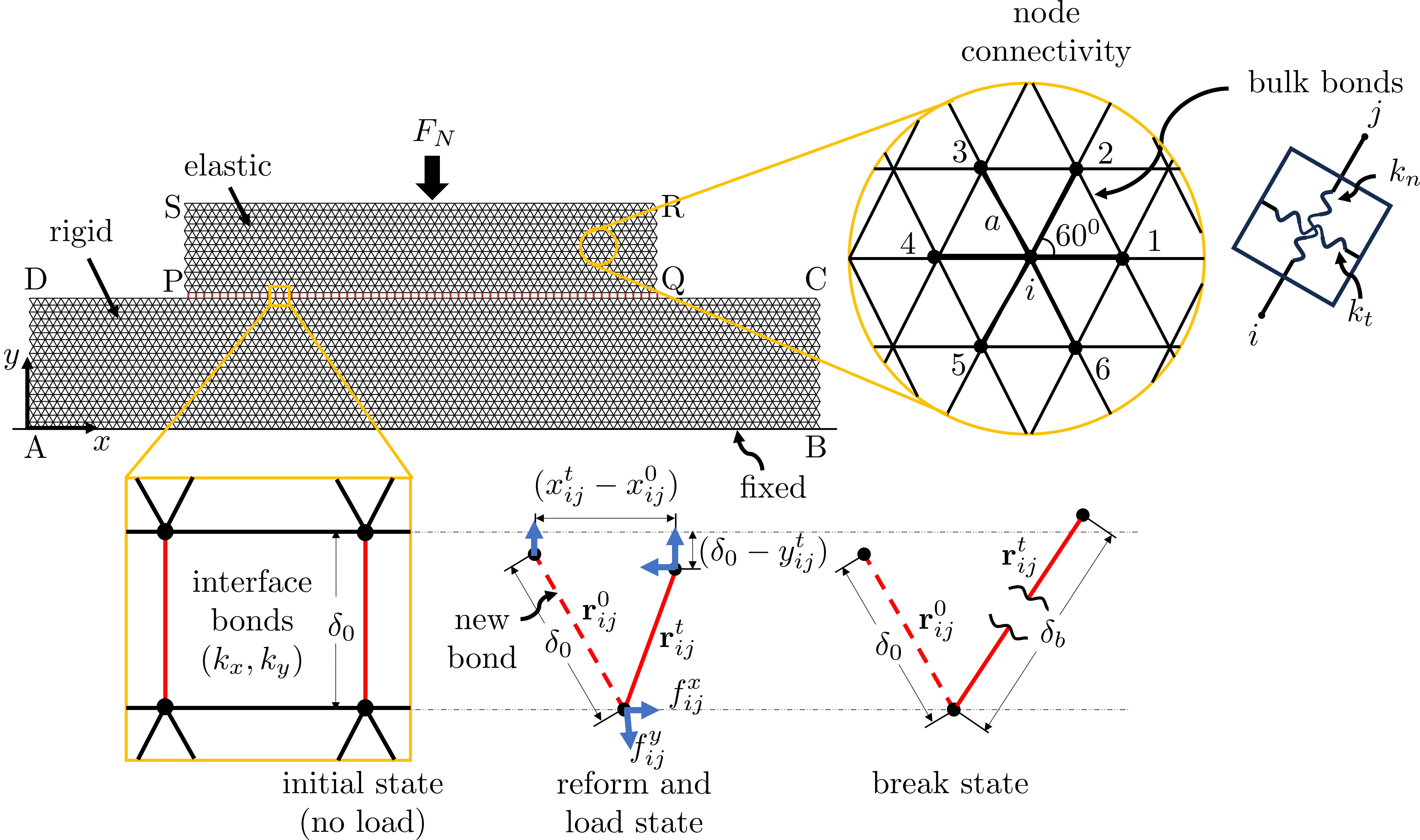}
	\caption{Schematic of lattice network representing two elastic bodies forming frictional bonds highlighted in red. The magnified view on the right displays the node connectivity and bond stiffness within the elastic body. Bottom panel shows loading behaviour and different states of interface bonds.}		
	\label{fig:schematic}
\end{figure}

The configuration we investigate is conceptually simple and is comprised of an elastic block slid over another stationary, comparatively rigid, block at constant speed $v_0$. Both blocks are represented by a 2D triangular network of load bearing bonds, commonly termed a Born network \cite{martin2000dynamic}, see Fig.~\ref{fig:schematic}. The network is 6-coordinated and has lattice spacing $a$ between adjacent nodes $i,j$. The positions of these nodes are denoted $\mathbf{r}_i, \mathbf{r}_j$ respectively, and the corresponding bond vector is $\hat{\mathbf{r}}_{ij}^0 =  \mathbf{r}_i -  \mathbf{r}_j$. Two such nodes are connected by springs that resist deformation perpendicular to and along $\hat{\mathbf{r}}_{ij}$, with spring constants $k_n$ and $k_t$ respectively, see top right inset to Fig.~\ref{fig:schematic}. Such bonds are termed \lq bulk bonds\rq\ to distinguish them from interface bonds between the two bodies.

The equation of motion $\mathbf{r}_i$ of node $i$ is
\begin{align}
	\frac{\partial^2 \mathbf{r}_i}{\partial t^2} = (k_n -k_t)\sum_{j=1}^{6}[(\mathbf{r}_{j}-\mathbf{r}_i - a\mathbf{\hat{r}}_{ij}^0).\hat{\mathbf{r}}_{ij}^0]\hat{\mathbf{r}}_{ij}^0 + k_t\sum_{j=1}^{6}(\mathbf{r}_{j} - \mathbf{r}_i - a\mathbf{\hat{r}}_{ij}^0)\equiv \mathbf{F}_{i}\label{eq:born}
\end{align}

In order to set the constants $k_t, k_n, a$, we take the continuum limit of this equation and compare with the corresponding shear and longitudinal wave speeds $c_s, c_l$ in Lam\'e equations of elastodynamics to get
\begin{align}
	k_n = \frac{c_l^2 - c_s^2/3}{a^2} \hspace{0.5cm}\text{and}\hspace{0.5cm} k_t = \frac{c_s^2 - c_l^2/3}{a^2} \label{eq:stiffness}
\end{align}

We found that this discretized model is numerically dispersive (wave speed depends on wavelength) for $\lambda/a < 7$. All of the results reported in this manuscript are beyond this limit. 

In addition to the forces between neighbouring lattice points (RHS of Eq.~\ref{eq:born}), we must also account for external loading and interface forces between the bodies. Nodes along the lines DC and PQ in Fig.~\ref{fig:schematic} interact with each other via interface bonds, see inset in bottom left, that are envisaged as adhesive junctions. The corresponding bond direction is $\mathbf{r}_{ij} = x_{ij}\hat{\mathbf{i}} + y_{ij}\hat{\mathbf{j}}$, with components $x_{ij}$ and $y_{ij}$ along $x$ and $y$ directions, respectively. At the initial state, the vertical separation $y_{ij}^0 = \delta_0$ between two nearest facing nodes on each body, see  bottom panel of Fig. \ref{fig:schematic}. This $\delta_0$ can be regarded as the limit to which the two blocks can be brought closer without exerting any normal force. When a net normal load $F_N$ is applied to the top body, interface bonds resist vertical compression $(\delta_0-y_{ij})$ and horizontal stretch by exerting corresponding forces $f_{ij}^y, f_{ij}^x$ 
\begin{align}
f_{ij}^y = k_y(y_{ij} - \delta_0) \quad\quad f_{ij}^x = k_x(x_{ij} - x_{ij}^0)
\end{align}
where $k_x, k_y$ represent the corresponding stiffness values of the interface spring, see bottom panel of Fig.~\ref{fig:schematic}. This is the characteristic of a single interface bond. As a result of loading, this bond gradually deforms until its length $\delta_{ij}$ reaches a threshold value $\delta_b$ ($>\delta_0$). At that instant, the bond breaks, allowing nodes to slide freely, see right bottom panel in Fig.~\ref{fig:schematic}. An interface bond may reform when two facing free nodes come close enough so that $\delta_{ij}^{\text{new}} \leq \delta_0$. This reformation is solely a function of the physical proximity for adhesion to be active between the objects. The total force on the interface nodes (along lines PQ and DC) is given by
\begin{equation}
  \mathbf{F}_f = f_{ij}^x\mathbf{\hat{i}} + f_{ij}^y\mathbf{\hat{j}}
\end{equation}

After incorporating numerical damping in Eq.~\ref{eq:born} and accounting for both $\mathbf{F}_f$ above and any external forces $\mathbf{F}_e$, the final equation of motion can be written as
\begin{align}
\frac{\partial^2 \mathbf{r}_i}{\partial t^2} =  \mathbf{F}_{i} + \mathbf{F_e} + \mathbf{F}_f - \zeta\frac{\partial \mathbf{r}_i}{\partial t} \label{eq:finaleq}
\end{align} 

This equation is solved using a time-explicit Verlet scheme
\begin{equation}
	\mathbf{r}_i(t+1) = 2\mathbf{r}_i(t) - \mathbf{r}_i(t-1) + \Delta t\mathbf{\ddot{r}}_i(t)
\end{equation}
where, the time increment $\Delta t$ must satisfy $\Delta t < a/c_l$ to ensure stability \cite{del2002wave,braun2014new}.

The initial configuration after bringing the two bodies in contact is shown in Fig.~\ref{fig:schematic}. The top and bottom blocks have sizes $10 \times 100$ and $20\times 200$, respectively. These exact sizes were confirmed to have little bearing on the results reported later.  The top block is chosen to be compliant, with $k_n, k_t$ chosen as per Eq.~\ref{eq:stiffness} to ensure that it has Poisson ratio of 0.2. The bottom lattice is made comparatively rigid by setting $k_n^2 = 100, k_t^2 = 0$. The applied normal load is normalized ($F_N = 1$) and distributed uniformly over all the nodes on the boundary SR of the upper body. For the bottom body, nodes along the line AB are all fixed to have zero displacement. Note that the distance and velocity are normalized by the initial contact length $l$ (here 100$a$) and Rayleigh wave speed $c_r$, respectively. The interface bond properties are set to be $k_x = 0.1, k_y = 0.3$ with $\delta = 0.5$. These values correspond to strong, adhesion at the interface. The nature of the results are unchanged for values close to the ones we have chosen. We also set $\zeta = 0.1$ to ensure convergence. 

Once the two bodies are in equilibrium after $F_N$ is applied, we apply a constant velocity boundary condition $v_0$ on the top body to initiate sliding. This boundary condition is applied in three different ways, corresponding to three different study cases---as shear on the top face SR (termed $T_0$), as a normal pushing stress on the left face SP (termed $T_{-}$) and as a normal pulling stress on the right face RQ (termed $T_{+}$). Numerical damping constant $\zeta = 0.1$ for all the simulations is found to be reasonable for damping nodes' oscillations.


\section{Results}\label{sec:results}
We present the results of our investigations of the model described in Sec.~\ref{sec:model}. The central finding is that each of the loading cases $T_{+}, T_{-}, T_{0}$ results in a different type of slip initiation. Note that the colour scheme used in this section is as follows: blue, orange, green and brown colours represent first, second, third and fourth temporal fronts, respectively, unless specified otherwise.

\subsection{Pushing induces forward moving pulse-like rupture}\label{sec:push}
We first consider the $T_{-}$ case where a constant pushing velocity $v_0$ is applied on the left face of the sliding body, see Fig.~\ref{fig:left_fronts}(a). Slip at the interface occurs via the periodic nucleation and propagation of rupture pulses starting from the left end of the block $x/l=0$. A single rupture front nucleates in the form of a broken interface bond at this end, and begins to grow by breaking adjacent bonds in the sliding direction towards $x/l=1$. Simultaneously, bond reformation occurs at the trailing end so that the rupture has a finite width, and represents a propagating pulse. The inset to panel (a) in Fig.~\ref{fig:left_fronts} shows two such pulses, along with the corresponding pulse tip (marked $x_{\text{tip}}$). 

\subsubsection{Propagation speed and interface bond states}

\begin{figure}[ht!]
	\centering	
	\includegraphics[width=\linewidth]{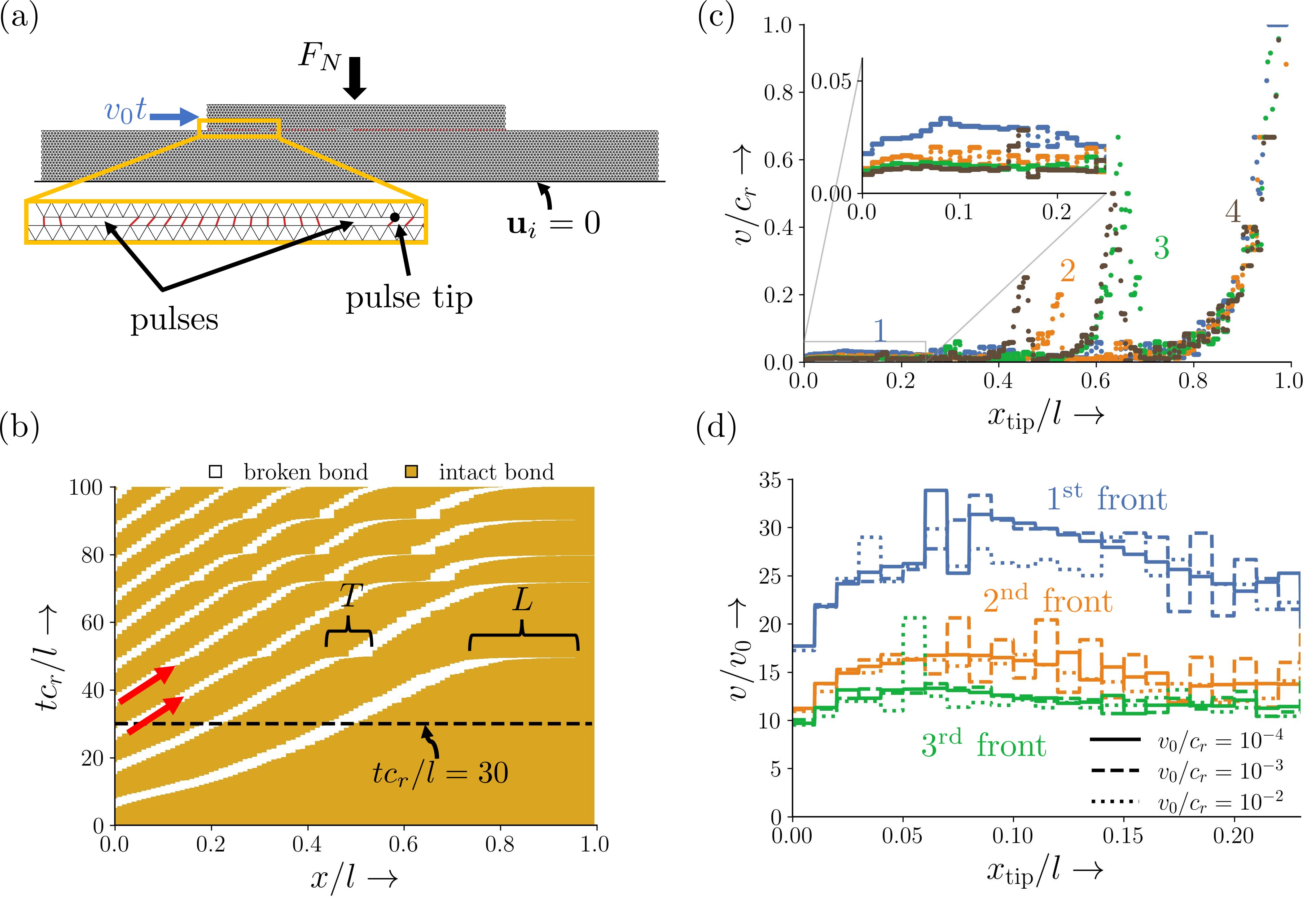}
	\caption{(a) Pulse-like cracks at $tc_r/l=30$ for pushing scenario with sliding speed $v_0/c_r = 10^{-3}$. The bottom and top lattice sizes are $20\times 300$ and $10\times 100$ with following parameters: $\nu = 0.2$, $k_x=0.1$, $k_y = 0.3$, $F_N = 1$, $\delta_0 = 0.5$, and $\delta_b = 0.7$. (b) Space-time plot of intact bonds at the interface. Red arrows indicate the pulse propagating direction. The dashed line corresponds to the time instance for which the interface snapshot is shown in (a). (c) Corresponding front velocity as it traverses the interface. Distinct colours of the curves depict 1 to 4 fronts, numbered by respective colours. (d) Comparison of front velocity for different sliding speeds for the same configuration.}
	\label{fig:left_fronts}
      \end{figure}

To quantify pulse propagation dynamics, a space-time diagram is constructed by stacking interface bond states along $x/l$ as a function of time ($t c_r/l$), see Fig.~\ref{fig:left_fronts}(b). Note that broken (intact) bonds are represented by white (yellow) colours. Pulses appear in this figure as white streaks corresponding to locally broken bonds at the interface. They begin at the left end ($x/l=0$) and propagate towards the right ($x/l=1$) with initial velocity much lesser than the Rayleigh wave speed (see red arrows); ten such pulses are seen in this figure. Several features may be inferred at once, in addition to their propagation direction. Firstly, the pulses all have constant width, as can be established by taking a constant time section (horizontal line) in this figure, see dashed line at $t c_r/l=30$. Data corresponding to these widths are provided in Fig.~\ref{fig:app_pw} of the appendix. Secondly, each pulse propagates at nearly constant speed through the interface (constant slope of white curves) until it reaches the right end, when it begins to accelerate. The first or leading pulse, marked $L$, begins to accelerate after $x/l=0.7$, causing the trailing pulses, marked $T$, to also follow suit. This pattern recurs when subsequent pulses approach the right end of the interface. Finally, the frequency at which pulses are nucleated at $x/l=0$, obtained using a vertical line in the figure, is nearly constant---motion of the interface occurs only via the nucleation and propagation of such pulses. The amount of interface slip caused by a single pulse will be quantified for all three loading conditions at the end of Sec.~\ref{sec:results}.

\subsubsection{Pulse speed variation across the interface}
Instantaneous speeds can be computed from the slope of the space-time diagram in Fig.~\ref{fig:left_fronts}(b) as the pulse traverses the entire interface. This is done in co-moving coordinates by first locating the tip $x_{\text{tip}}$ of the pulse (first broken bond) and determining the speed $v$ at this location. The result for four successive pulses (coloured blue, orange, green, brown and marked 1,2,3,4 respectively) is summarized in Fig.~\ref{fig:left_fronts}(c). The vertical axis in this figure is normalized by the Rayleigh wave speed $c_r$. The fact that all these pulses propagate at speeds much lesser than the Rayleigh wave speed is clear from the inset to this panel. An additional feature is that as the leading pulse (blue) accelerates, \emph{viz.} when its tip reaches $\sim 0.8l$, the other trailing pulses accelerate as well, see the coincident increase in $v/c_r$ for all four pulses. The same is observed when each of these pulses reach the right end of the contact. Apart from this end of the contact, the pulse velocity $v$ appears to have little to do with the Rayleigh speed $c_r$---all four pulses travel at nearly the same speed as seen in the inset to panel (c) of Fig.~\ref{fig:left_fronts}.

The only other velocity scale in the problem is the applied $v_0$---in order to test the dependence of the pulse velocity $v$ on $v_0$, we collated data for three different sliding speeds $v_0/c_r = 10^{-2}, 10^{-3}, 10^{-4}$, see dotted, dashed and solid lines, respectively, in panel (d) of Fig.~\ref{fig:left_fronts}. It is clear that the first, second and third pulses in each case travel with a speed $v \propto v_0$ since the corresponding coloured curves appear nearly on top of each other. Consequently, the pulses move only as fast as the remote pulling velocity until they accelerate to $c_r$ at the right end of the contact. 

\subsubsection{Stress field dynamics of pulse nucleation and propagation}
\begin{figure}[ht!]
	\centering	
	\includegraphics[width=\linewidth]{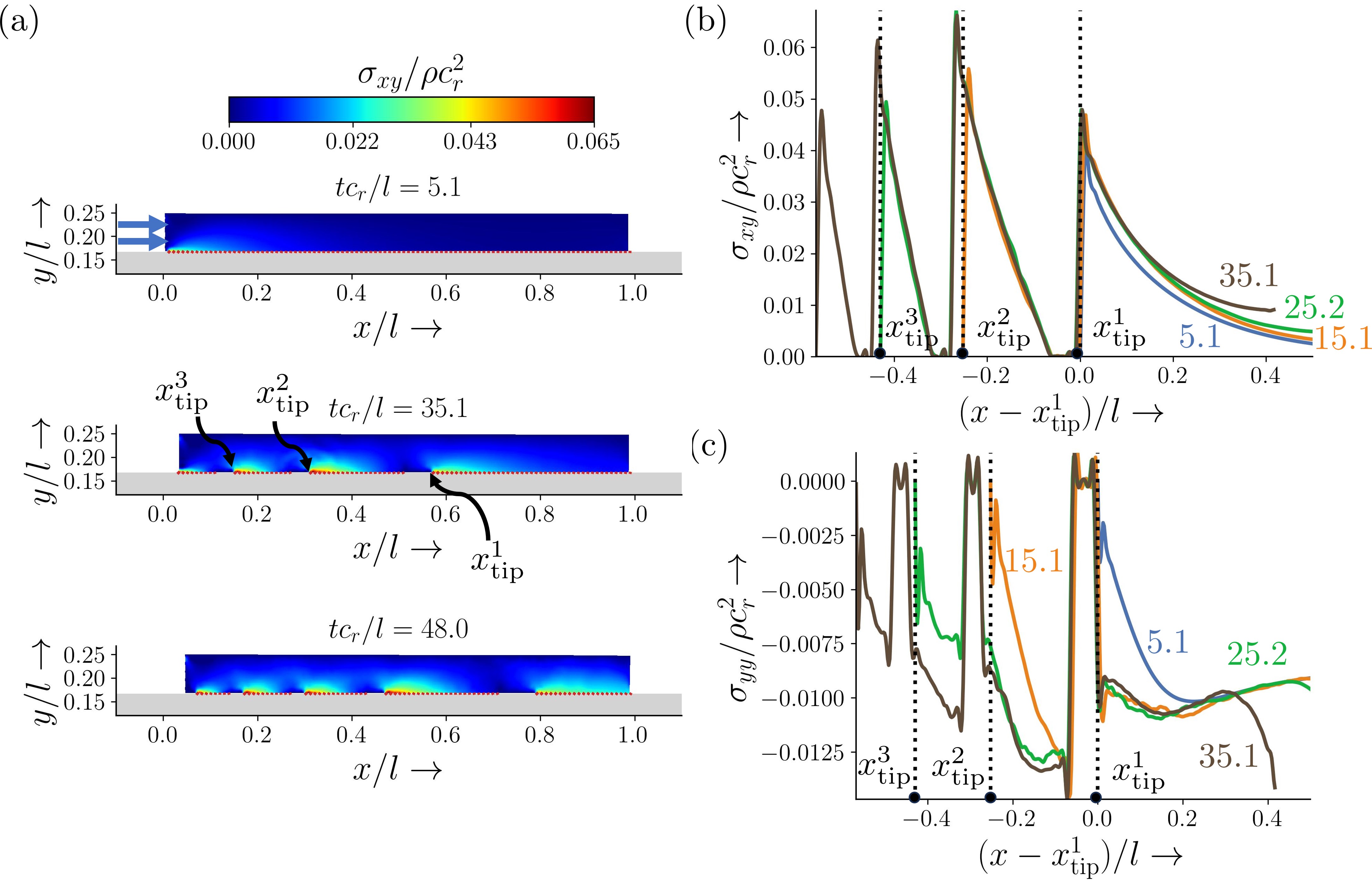}
	\caption{(a) Stress field $\sigma_{xy}$ together with the interface bonds (red colors) are shown at three different time instances for pushing scenario. The sliding speed $v_0$, normal load $F_N$, and bond parameters are identical to Figure \ref{fig:left_fronts}. (b) and (c) Extracted shear $\sigma_{xy}$ and normal $\sigma_{yy}$ at the interface in front of the first ($x_{tip}^1$), second ($x_{tip}^2$), and third ($x_{tip}^3$) front tip. The coloured number indicates the corresponding time instances $tc_r/l$ of 5.1 (blue), 15.1 (orange), 25.2 (green), and 35.1 (brown).}		
	\label{fig:left_stress}
\end{figure}

In addition to the kinematics of pulse propagation and interface slip, we now examine the evolution of stresses in the elastic body prior to, during and after pulse propagation. For this $T_{-}$ case, the shear stress $\sigma_{xy}$ in the elastic body is shown as a colour map in the sequence reproduced in panel (a) of Fig.~\ref{fig:left_stress}. Note that the stress is reported non-dimensionlized by $\rho c_r^2$ for three different times corresponding to different pulse nucleation events. When $tc_r/l=5.1$, the first interface bond is about to break, the developed stress field clearly exhibits a local maximum at the left (rear) edge $x/l=0$. Once a pulse is nucleated, the normal stress decreases rapidly in its wake; this is then followed by another similar stress build-up at the left edge and another nucleation event ensues. Subsequently, the stress field arising from the interaction of three pulses ($x_{\text{tip}}^1$, $x_{\text{tip}}^2$, $x_{\text{tip}}^3$) is depicted for time $tc_r/l= 35.1$; another new pulse is about to simultaneously nucleate at $x/l=0$. Finally, the stress field is shown at time $tc_r/l = 48.0$ when the leading pulse interacts with the other boundary. This interaction amplifies the stress field ahead of all pulses, causing them to accelerate.

We plot the shear $\sigma_{xy}$ and normal $\sigma_{yy}$ stresses along the interface with respect to the leading front tip at four different times in Fig.~\ref{fig:left_stress}(b) and (c), respectively. The timestamps correspond to four distinct nucleation events at the left edge of the contact. These distinct times pertain to the instant when a new pulse nucleates at the left (rear) edge---the first one at $tc_r/l = 5.1$, second at $tc_r/l = 15.1$, and so forth. As before, the vertical dashed lines labeled $x_{\rm{tip}}^i$ correspond to the position of $i^{th}$ pulse tip. Note also that the horizontal axes in both plots are centered around $x_{\text{tip}}^1$. Two key observations emerge from these stress variations: Firstly, the shear stress peak at the onset of pulse nucleation is quite high; as is the normal stress. This implies that pulse nucleation is influenced by both these components, quite reminiscent of mixed-mode interface fracture. Secondly, the stress peak, in both shear and normal stress components, at the tip of the pulse remains constant during propagation, explaining the steady pulse speed and the consistent spacing between pulses. Remarkably, we also found that the shear stress in the vicinity of the pulse tip appears to exhibit a square-root singular behaviour, characteristic of mode II fracture (see appendix~\ref{sec:app_singular}).

In summary, the onset of the rupture front depends on both stress components when the block undergoes pushing. The localised stress field along the interface, with a relatively steady stress peak level, develops during its propagation. This results in slow front propagation, providing sufficient time for the bonds to readhere.

\subsection{Pulling induces opposite moving pulse-like rupture}\label{sec:pull}
\begin{figure}[ht!]
	\centering	
	\includegraphics[width=\linewidth]{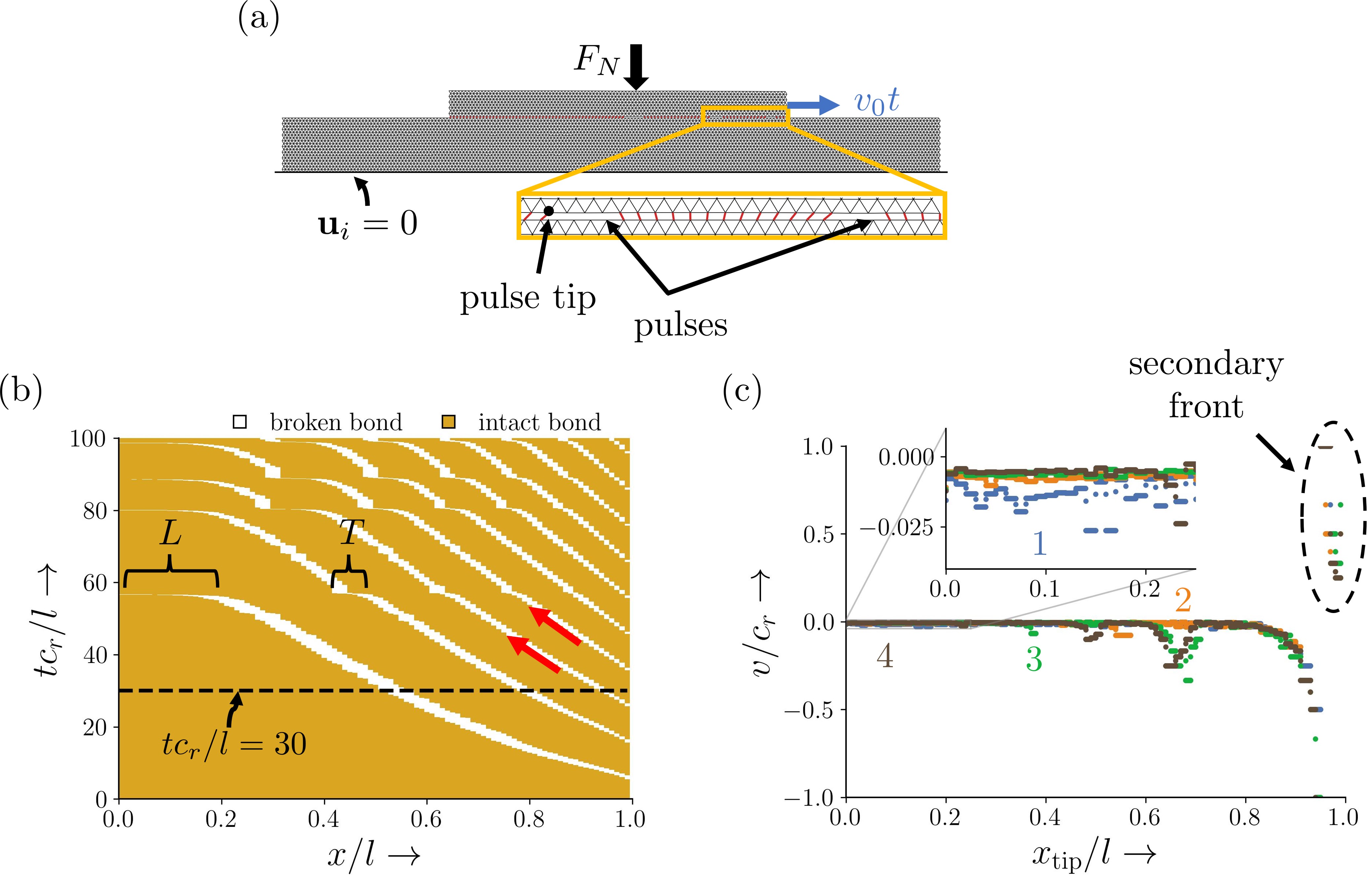}
	\caption{(a) Pulse-like cracks at $tc_r/l=30$ for pulling scenario with sliding speed $v_0/c_r = 10^{-3}$. The sliding speed $v_0$, normal load $F_N$, and bond parameters are identical to Figure \ref{fig:left_fronts}.  (b) Space-time plot of intact bonds at the interface. Red arrows indicate the pulse propagating direction. The dashed line corresponds to the time instance for which the interface snapshot is shown in (a). (c) Corresponding front velocity as it traverses the interface. Distinct colours of the curves depict 1 to 4 fronts, numbered by respective colours.}		
	\label{fig:right_fronts}
\end{figure}

The case of $\Tp$ is analogous to the $\Tm$ case just discussed, with a pull being applied in the form of a constant velocity $v_0$ on the leading face, \emph{RQ} in Fig.~\ref{fig:schematic}. Similar pulses are observed, but they now nucleate at the leading face and propagate towards the rear, or in a direction opposite to the imposed motion, see Fig.~\ref{fig:right_fronts}. Panel (a) of this figure shows the interface bonds and the corresponding pulse tip, see inset.

\subsubsection{Propagation speed and interface bond states}

A corresponding space-time diagram for this $\Tp$ case is reproduced in Fig.~\ref{fig:right_fronts}(b). Here again, yellow (white) colours correspond to intact (broken) bonds. The pulse is seen as an oriented white band (see red arrows) that appears to have a constant propagation speed in the initial stages of propagation from $x/l=1$. The speeds are far lower than the corresponding Rayleigh wave speed. Completely analogously to the $\Tm$ case discussed before, these pulses accelerate when the approach the other end, here $x/l=0$. The leading pulse (marked $L$) first accelerates, as do the trailing (marked $T$) pulses simultaneously. By simple comparison, panels (b) in both Figs.~\ref{fig:right_fronts},~\ref{fig:left_fronts} appear to be identical mirror images of each other---consequently, all of our discussion there also applies to the $\Tp$ case, save for the change in direction of wave propagation.
      
\subsubsection{Pulse speed variation across the interface}

Just as we did for the $\Tm$ case, we extract velocity data from the space time diagram corresponding to four consecutive pulses at the interface, see panel (c) of Fig.~\ref{fig:right_fronts}. Note that the speed is represented as a signed quantity---negative values imply motion in the opposite direction to $v_0$. We see that all pulses travel at speeds much lower than the Rayleigh wave speed, with a scaling $v \sim v_0$ similar to the $\Tm$ case. Again, peaks in the curves occur as a consequence of acceleration of the leading pulse near the opposite boundary, causing acceleration of all trailing pulses.  In contrast to the $\Tm$ case, we now observe ocassional secondary fronts that appear to originate from the trailing edge $x/l \sim 0.1$  from the rear boundary, and move towards the existing pulse. Like for $\Tm$ case, it's evident that the velocities of these opposite moving pulses also linearly scale with the sliding velocity $v_0$ and this result is, thus, omitted here for brevity.

\subsubsection{Stress field dynamics of pulse nucleation and propagation}

\begin{figure}[ht!]
	\centering	
	\includegraphics[width=\linewidth]{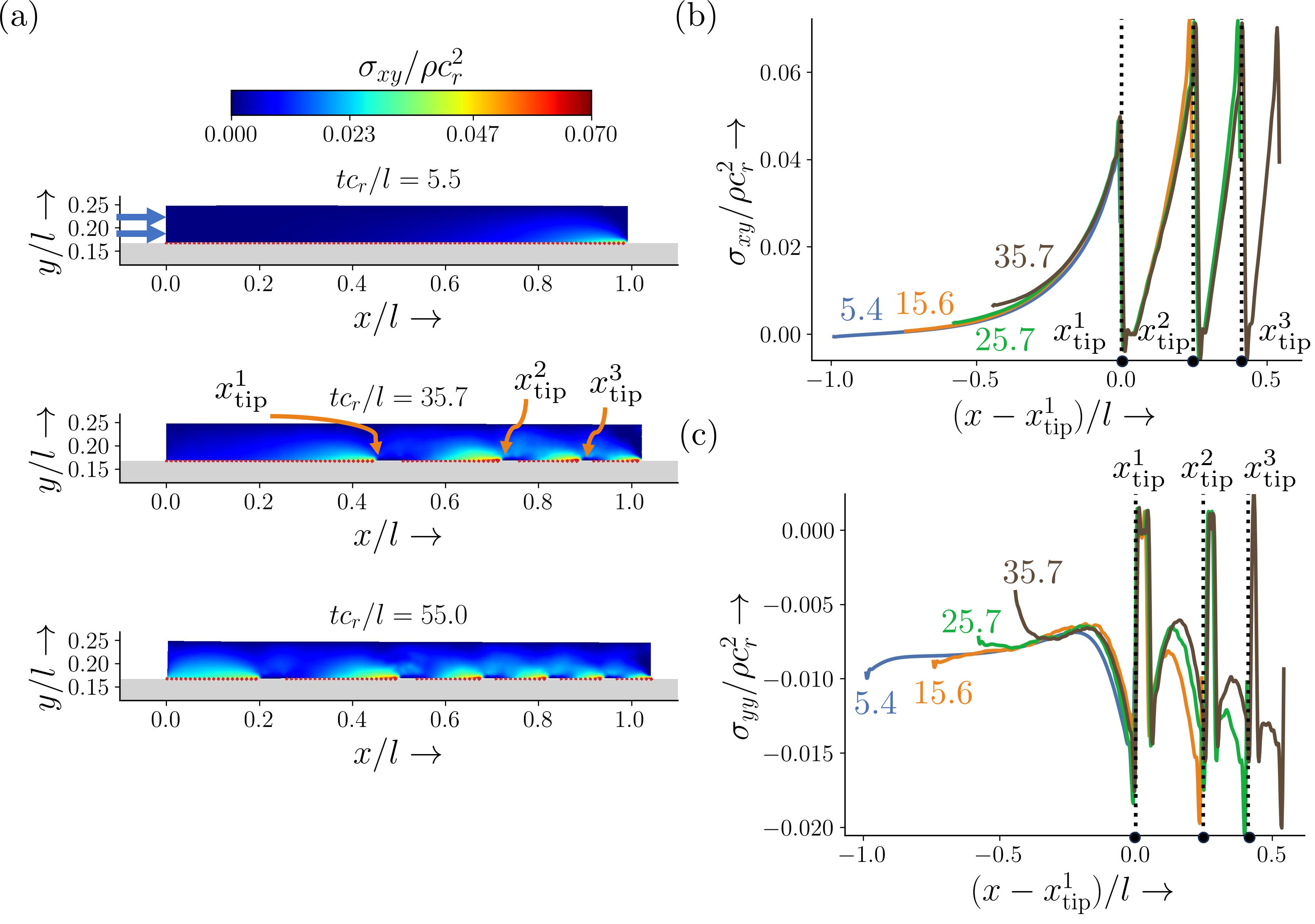}
	\caption{(a) Stress field $\sigma_{xy}$ together with the interface bonds (red colors) are shown at three different time instances for puling scenario. The sliding speed $v_0$, normal load $F_N$, and bond parameters are identical to Figure \ref{fig:left_fronts}. (b) and (c) Extracted shear $\sigma_{xy}$ and normal $\sigma_{yy}$ at the interface in front of the first ($x_{tip}^1$), second ($x_{tip}^2$), and third ($x_{tip}^3$) front tip. The coloured number indicates the corresponding time instances $tc_r/l$ of 5.4 (blue), 15.5 (orange), 25.7 (green), and 35.7 (brown).}		
	\label{fig:rightstress}
      \end{figure}

We now plot the interface stresses $\sigma_{xy}/\rho c_r^2$ as a function of time, see panel (a) in Fig.~\ref{fig:rightstress}. Again, the situation is analogous to the $\Tm$ case---a stress peak occurs just prior to pulse nucleation, followed by relaxation. Several pulses are seen subseqently, their tips are marked individually for $tc_r/l = 35.7$ when the fourth pulse is about to nucleate. At $tc_r/l = 55$, the stress field of the leading pulse begins to interact with the rear edge of the body, at which point it begins to accelerate. 

We next present a comparison between interface shear $\sigma_{xy}$ and normal $\sigma_{yy}$ stresses, plotted with respect to the location of the first pulse tip ($x_{\text{tip}}^1$), see panels (b) and (c) of Fig. \ref{fig:right_fronts}. Four different times are denoted, corresponding to the stress fields in panel (a). The shear stress remains almost constant post nucleation while the normal stress has a peak at tip. The latter reduces the pulse speed by 50\%, see the discussion in appendix~\ref{sec:app_speed}.

\subsubsection{Tensile vs. compressive nature of $\Tp, \Tm$ pulses}
\begin{figure}[ht!]
	\centering	
	\includegraphics[width=0.95\linewidth]{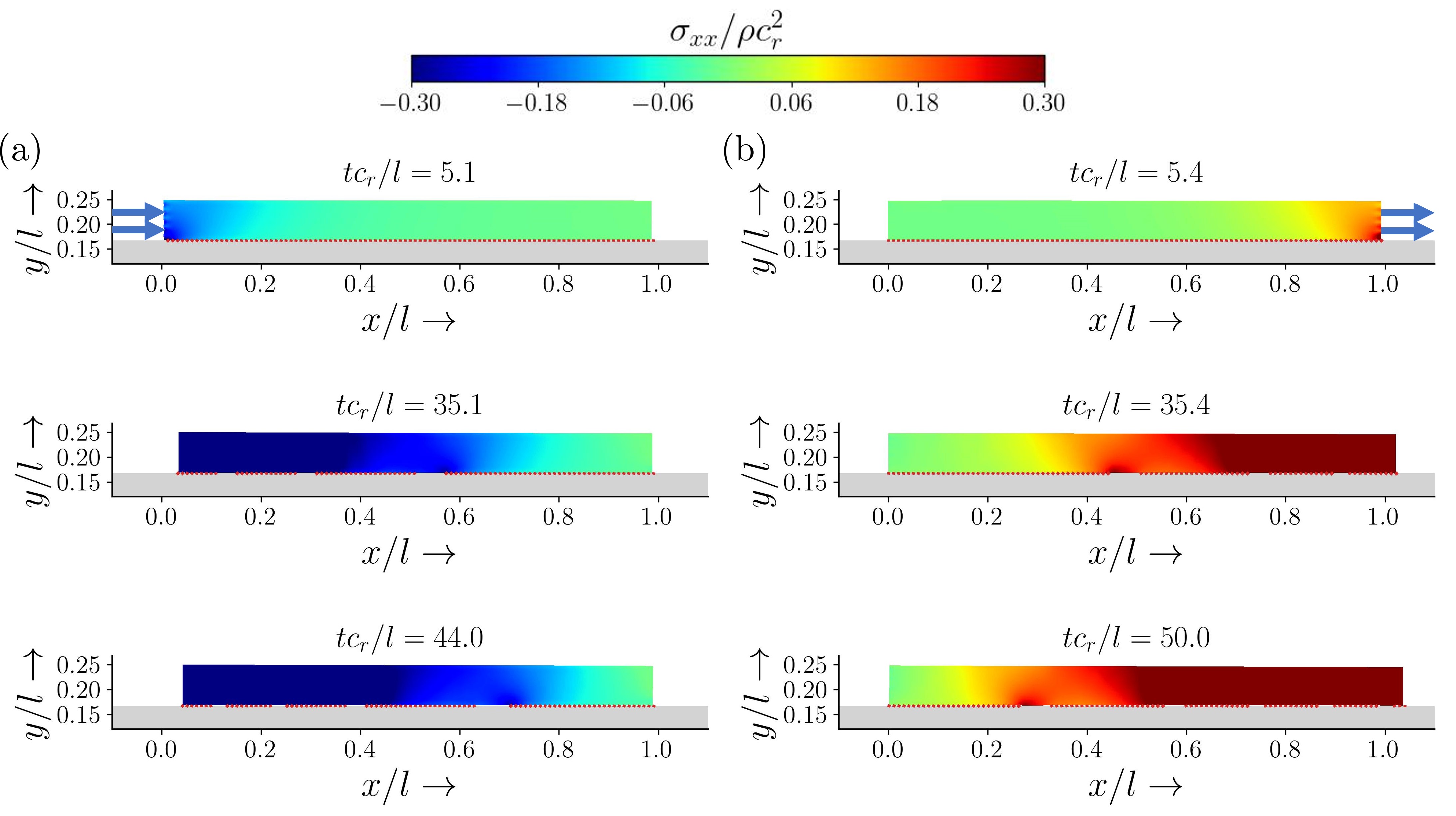}
	\caption{Stress field $\sigma_{xx}$ along with the interface bonds (red colors) at three different time instances for the (a) pushing and (b) pulling configurations. The driving velocity $v_0$, normal load, and bond parameters are same as Figure \ref{fig:left_fronts}.}		
	\label{fig:left_right_sxx}
\end{figure}

The two cases $\Tm, \Tp$ differ only in the direction of wave propagation, a property that may be attributed to the tensile or compressive nature of the strains/ stresses. To illustrate this, let the horizontal displacement $u_x$ be a function of the form $u_x(x - v t)$ where $v$ is the wave speed. The longitudinal strain $\epsilon_{xx} \sim \frac{\partial u_x}{\partial x}$ is hence:
\begin{equation}
  \epsilon_{xx} = \frac{\partial u_x}{\partial x} = -v^{-1}\frac{\partial u_x}{\partial t} \sim v^{-1} \Delta x/T 
\end{equation}
where $\Delta x$ is the amount of slip induced by a single wave (quantified in a later section) and $T$ is the propagation time (vertical coordinate in the space-time diagram). For $\Tm$, the observed $v$ is positive, so that $\epsilon_{xx}$ is compressive. The converse holds for $\Tp$, the longitudinal strain component is tensile. We establish this from the simulations by plotting the corresponding $\sigma_{xx}$ fields for both $\Tm$ and $\Tp$ cases, see Fig.~\ref{fig:left_right_sxx}. The figure depicts three distinct time instances, corresponding to a new pulse generation event, of the $\sigma_{xx}$ field, along with showcasing the interface bonds in red. Reattachment of interface bonds in either case is facilitated by these $\sigma_{xx}$ values, even though both $\sigma_{yy}$ and $\sigma_{xy}$ are responsible for their nucleation. This reattachment is perhaps the primary reason why these ruptures propagate in the form of localized pulses. 

\subsection{Crack-like rupture mode in sliding}\label{sec:slide}
Finally, we now discuss the $\To$ case, with the loading applied on the top face ($SR$ in Fig.~\ref{fig:schematic}). Here the interface dynamics differs significantly from the $\Tm, \Tp$ cases, as summarized in Fig.~\ref{fig:top_fronts}. Rupture fronts nucleate at the left edge of the body (see inset to panel (a)), just as in the $\Tm$ case. However, the rupture propagates very fast with $v_0 \sim c_r$ so that reattachment does not occur. Interface motion thus now occurs via these crack-like rupture fronts.

\subsubsection{Propagation speed and interface bond states}

\begin{figure}[ht!]
	\centering	
	\includegraphics[width=\linewidth]{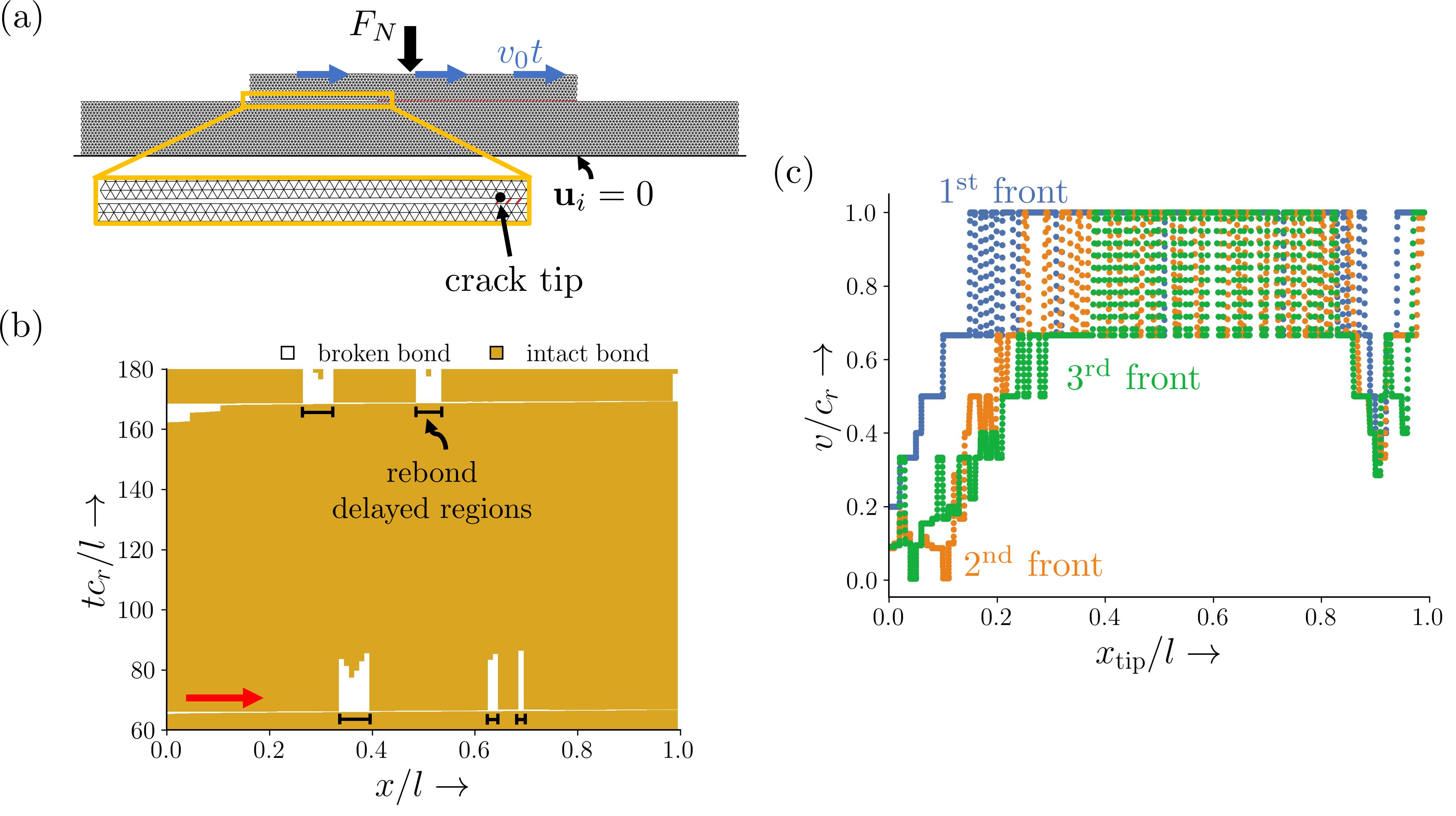}
	\caption{(a) Rupture front at $tc_r/l=66$ for sliding scenario with sliding speed $v_0/c_r = 10^{-4}$, and other parameters identical to Figure \ref{fig:left_fronts}.  (b) Space-time plot of intact bonds at the interface. Read arrow indicate the propagating direction. (c) Corresponding front velocity as it traverses the interface. Distinct colors of the curves depict 1 to 3 fronts, annotated by respective colors. }		
	\label{fig:top_fronts}
\end{figure}

The corresponding space-time diagram for the $\To$ case is reproduced in Fig.~\ref{fig:top_fronts}(b). Just as before, broken (intact) bonds are coloured white (yellow). The first rupture front nucleates at the left end $x/l=0$. The propagation direction is shown (red arrow) but now the motion appears instantaneous. The small slope in this figure implies that the rupture front has already reached the Rayleigh wave speed. Reattachment of interface bonds in the wake of the rupture does occur, however, the timescales are such that this process starts after the rupture has propagated the entire length of the interface. Reattachment is also non-homogeneous, with some regions remaining unbonded for a finite amount of time within the contact. These \lq rebond delayed\rq\ regions, marked in panel (b) continue to remain detacheed for a finite amount of time, and appear as short vertical bands in the space-time diagram. The caue for this is that interface bond reformation relies on the proximity of interface nodes reaching critical separation $\delta_0$. This in turn depends on the elastic relaxation of these nodes, leading to non-uniform occurrences of bond renewal events.

\subsubsection{Pulse speed and  stress dynamics}

The rupture front velocities are reproduced in panel (c) of Fig.~\ref{fig:top_fronts}. For three different fronts, coloured blue, orange and green, it is clear at once that the fronts quickly accelerate to $v \sim c_r$. The apparent fluctuation between $0.7c_r$ and $c_r$ in this panel is an artifact arising from the discreteness of the lattice. All temporally sequential fronts reach the Rayleigh wave speed quickly, even as $x_{\text{tip}}/l \sim 0.2$. It is noteworthy that fronts cannot propagate at speeds faster than this---we do not observe any super-shear ruptures.

\subsubsection{Stress field dynamics of pulse nucleation and propagation}

\begin{figure}[ht!]
	\centering	
	\includegraphics[width=\linewidth]{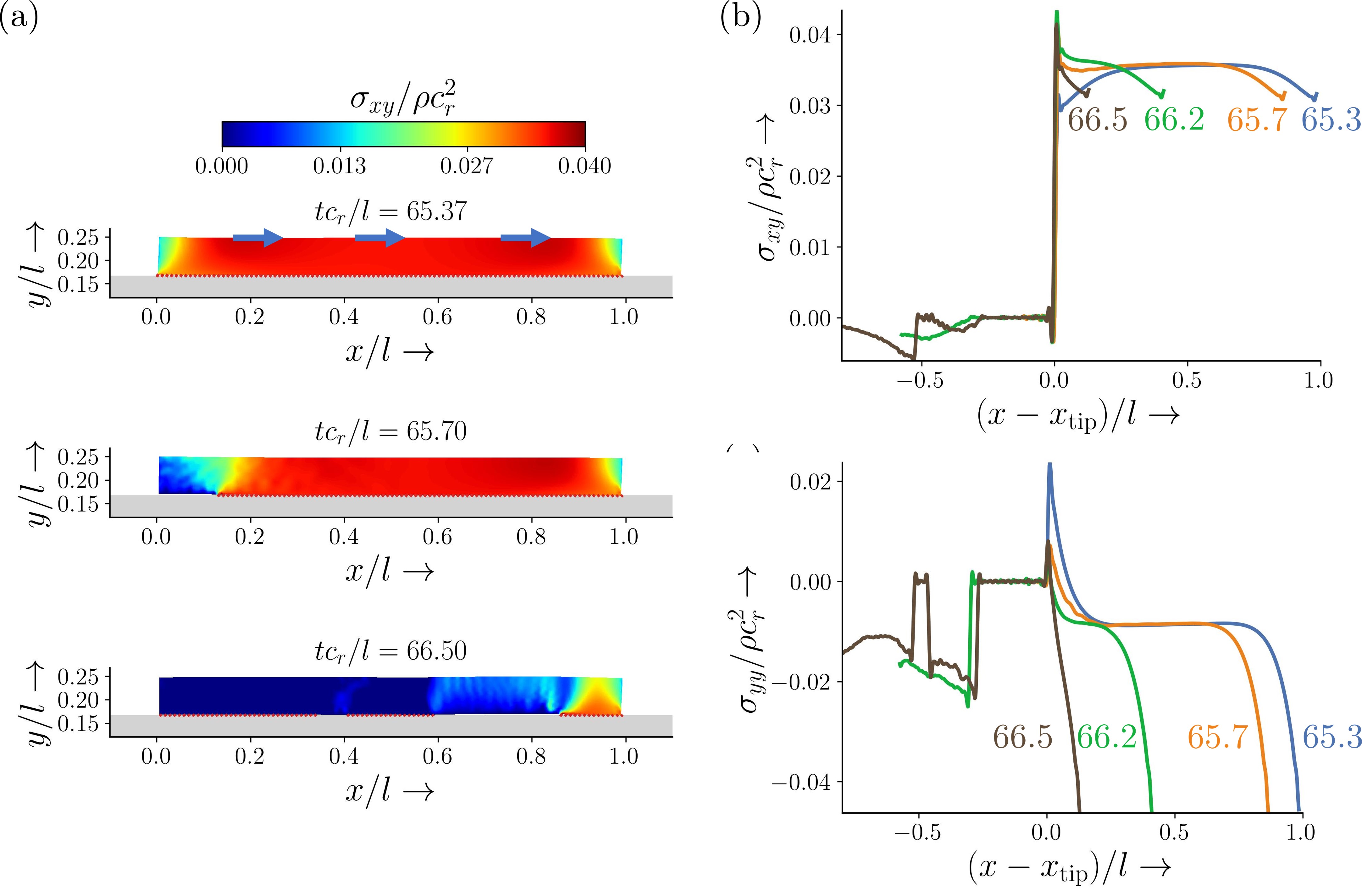}
	\caption{(a) Stress field $\sigma_{xy}$ together with the interface bonds (red colours) at three different time instances for top sliding cases with driving speed $v_0/c_r = 10^{-4}$, and other parameters identical to Figure \ref{fig:left_fronts}. (b) The extracted stress field $\sigma_{xy}$ at the interface ahead of the leading front at different time instances $tc_r/a$ of 65.3 (blue), 65.7 (orange), 66.2 (green), and 66.5 (brown). (c) Space-time plot of intact bonds with an inset image displaying the speed of the first (blue), second (orange), and third (green) fronts.}		
	\label{fig:top_stress}
\end{figure}

Just as in Fig. \ref{fig:left_stress} and \ref{fig:rightstress}, the spatial distribution of shear stress $\sigma_{xy} /\rho c_r^2$ for the $\To$ case, together with the interface bonds (highlighted in red), is depicted in Fig. \ref{fig:top_stress}(a). The picture is now signficantly different from $\Tm, \Tp$ cases. Firstly, the first breaking event occurs much later, at $tc_r/l = 65.37$. At this point, the shear stress is almost uniform throughout the body, except at the edges. This stands in stark contrast with the $\Tm, \Tp$ cases, where significant local stresses were observed at the contact edges at the nucleation onset. Secondly, the relaxation time for the stress is also extremely small post nucleation, vis-\'a-vis $\Tm, \Tp$ cases. The front propagates rapidly, relieving the elastic stress as shown for time $tc_r/l = 65.7$. Although rebonding occurs behind the rupture tip, these fast fronts rupture the entire interface before reformed bonds bear any load, see the $\sigma_{xy}$ field at $tc_r/l = 66.50$.

The corresponding interface shear $\sigma_{xy}$ and normal $\sigma_{yy}$ stresses are reproduced in Fig. \ref{fig:top_stress}(b) and (c), respectively. Prior to rupture initiation (blue curve), the finite geometry of the top lattice generates uneven shear stress at the contact interface, increasing from the edges to the central part of the contact interface. As bond breaking depends on local stretch, the additional contribution at the nucleation site (edge) comes from the normal stress having a positive maximum value. The highly stressed region ahead of the nucleated rupture ensures that this front accelerates quickly towards the Rayleigh wave by relieving energy from the stored elastic field. Notably, and unlike the $\Tp, \Tm$ cases, we observe stress peaks in both normal and shear components during rupture propagation. 

In summary, we deduce that the coupled effect of $\sigma_{xy}$ and $\sigma_{yy}$ stress components determines the nucleation site. Once nucleated, a fast front occurs in sliding because of the considerable shear stress accumulated along the interface. Consequently, a rupture propagates in a crack-like mode before a reformation process can affect its motion.

\subsection{How does the interface slide?}\label{sec:slip}
\begin{figure}[ht!]
	\centering	
	\includegraphics[width=\linewidth]{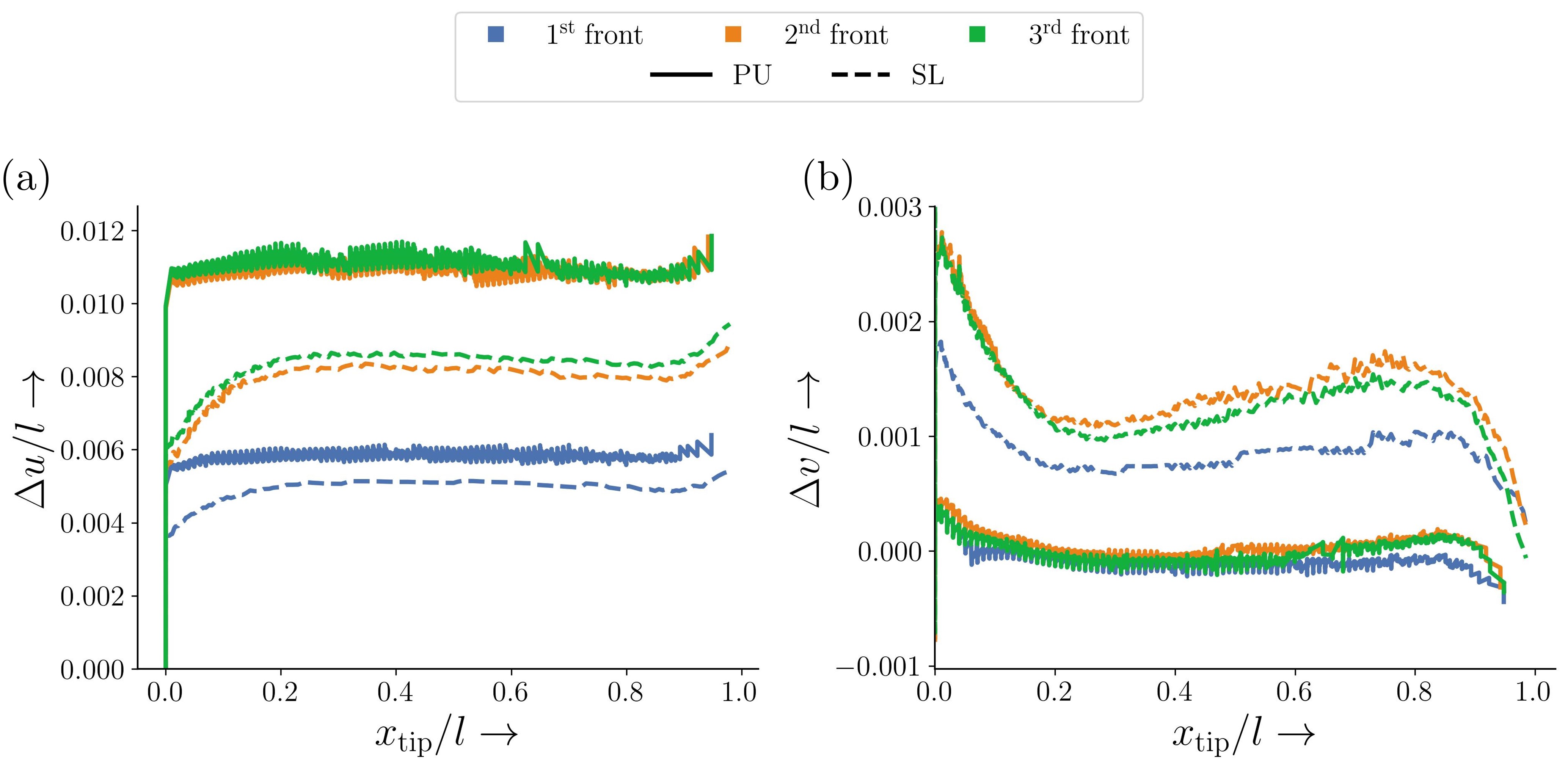}
	\caption{(a) Horizontal displacement $\Delta u$ and (b) vertical displacement $\Delta v$ occurring behind the leading edges of the first (blue), second (orange), third (green), and fourth (brown) propagating fronts as they traverse the interfaces under pushing (solid lines) and top sliding (dashed lines) loading conditions.}		
	\label{fig:movingslip}
\end{figure}

While interface ruptures propagate in the form of a crack or a pulse, the key property that relates these processes back to the original sliding is the amount of interface slip induced.  Since the slip behaviour in $\Tm,\Tp$ cases has reflection symmetry about the centre of the interface, we discuss only the $\Tm$ case. The horizontal interface slip $\Delta u$ associated with the $i^{th}$ front is obtained by subtracting from the total horizontal displacement $u$ of a comoving point the cumulative applied displacement $v_0t$ up to the instant the previous $(i-1)$ front nucleates. The $\Delta u$ vs. tip location is summarized in Fig.~\ref{fig:movingslip}(a). Data for three different fronts (blue, orange and green) are shown here in solid (dashed) lines corresponding to the $\Tm$ ($\To$) case. We note three features of the slip data. Firstly, local slip occurs only during pulse or rupture propagation at the interface. At other instances, i.e., between successive pulses, the interface remains stationary and there is no slip. Hence, it is clear that these fronts \lq mediate\rq\ the transition from static to dynamic friction. Secondly, $\Delta u$ for the first front, in both $\Tm, \To$ cases is generally larger than with subsequent pulses. This is most likely due to the effect of initial orientation of interface bonds, see also Sec.~\ref{sec:push}. Finally, each pulse or rupture front causes constant unit slip varying from $\sim 0.04 l$ to $0.011l$. Local slip is almost $25\%$ larger in the pulse ($\Tm$) case compared to the crack ($\To$) case. This is because $\Delta u$ is calculated from $v_0 t$ until the point of rupture nucleation. 

Analgoous behaviour is seen in the case of the vertical displacement $\Delta v$ due to a wave event, see Fig.~\ref{fig:movingslip}(b). Here, $\Delta v$ is defined as the vertical displacement $v$ from the initial equilibrium position (prior to $v_0$ application) of the same comoving point behind the front tip, as it travels across the interface. Clearly, $\Delta v >0$ at the left edge of the contact ($x/l = 0$) where nucleation occurs. This clearly indicates interface detachment that then subsequently relaxes to $\Delta v = 0$ in the wake of the wave at $x_{\text{tip}}/l = 1$. As expected, there is no net slip in the $y$ direction and the bodies \lq readhere\rq\ after wave passage. 

\begin{figure}[ht!]
	\centering	
	\includegraphics[width=\linewidth]{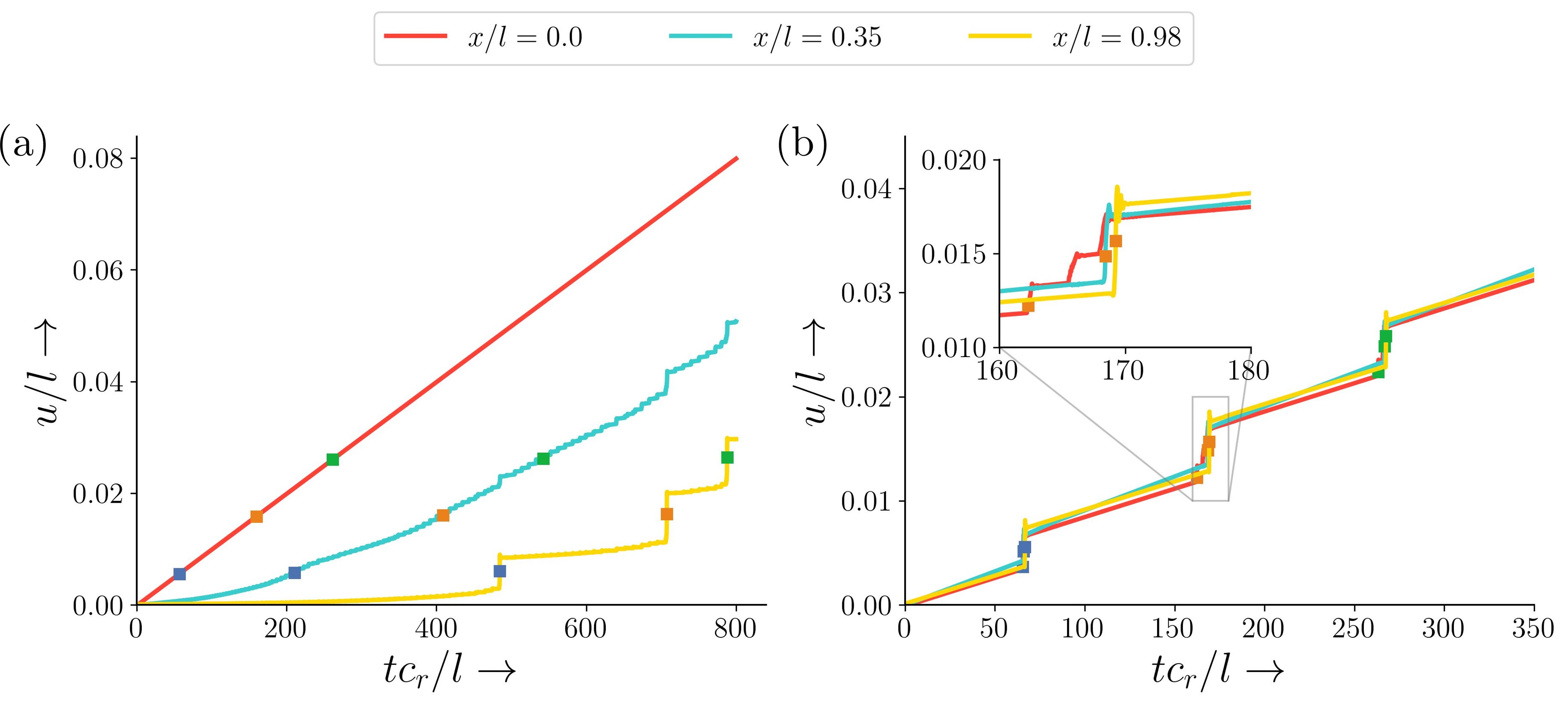}
	\caption{Temporal evolution of horizontal displacement $u$ at three distinct interface locations for both pushing and sliding scenarios, respectively. Color markers indicate the moment when the fronts pass these locations.}		
	\label{fig:uslip}
\end{figure}

These pulse/crack tip slip results can now be compared with the spatio-temporal slip variation at the interface as a whole, see Fig.~\ref{fig:uslip}. Panels (a) and (b) of this figure show the horizontal displacement $u$ for the $\Tm$ and $\To$ cases, respectively. The curves in these panels, coloured according to their location along the interface, show progress of interface slip with time. For the $\Tm$ case (panel (a)), the rear edge $x/l=0$ is driven with a constant velocity $v_0$ that is reflected in a line with constant slope (red). At other locations, $u$ progressively increases over time but is no longer linear---sudden jumps appear when a pulse traverses it. This effect is more pronounced as one approaches the right end $x/l = 1$; here slip occurs primarily during pulse passage, with the interface stationary in between. As earlier, blue, orange and green markers indicate successive pulses whose locations are superimposed on the curves in this panel. The fact that each pulse is responsible for constant slip is evident in that these markers all appear to be on corresponding horizontal lines. 

The situation is a bit more dramatic in the $\To$ case, see Fig.~\ref{fig:uslip}(b). Since the remote displacement is now applied far from the interface, we see all three locations appear to slip in the same manner with time. These curves now grow slowly with time until a fast rupture front arrives, at which point a finite additional slip is observed at all interface locations. The fact that pulses propagate at fast speed is clear from the inset---even though the markers appear to lie on top of each other, they are separated in time. The $u$ at any point is likewise cumulative slip $\Delta u$ carried by all prior fronts. This clear distinction between pulses and cracks is also evident from their slip velocities $\dot{u}$---with pulses having a lower and more gradual slip velocity than crack-like fronts with a sudden, sharp jump. Details of these velocity plots are discussed in appendix~\ref{sec:app_slipvel}.

\section{Discussion}\label{sec:discussion}
We now discuss the physical implications of our results and put some of our model observations in a broader context involving the transition from static to dynamic friction. We also briefly survey some existing theories for rupture/pulse mode selection in the context of our results. 

\subsection{Physical interpretation of the lattice network}
The presented lattice-based numerical model can capture two-dimensional dynamic displacements and stress fields for two finite-size elastic bodies in adhesive contact undergoing dynamic detachment and reattachment of macrocontacts governed by the interface deformation. An individual bond between two nodes in the Born network physically corresponds to an elastic interaction between two macroscopic segments of material in the bulk. Likewise, interface bonds can be pictured as macroscopic adhesive contact junctions. A bond deforms elastically and detaches at a finite strain level. The interface bond physics originate from the underlining microscopic mechanism such as attachment and detachment of macromolecular chains\cite{schallamach1963theory, filippov2004friction} in polymers.

The estimation of the bond stiffness for the interface can be done using the Hertz-Mindlin\cite{mindlin1949compliance} or JKR theory\cite{johnson1971surface,savkoor1977effect} for two perfectly adhesive elastic contacts under normal and tangential loading. The interacting distance $\delta_0$ may be set arbitrarily since only the difference in displacements from the initial value has physical significance. However, determining the breaking strain $\delta_b$ of the macro-contacts remains tricky. Regardless, one can predict the qualitative behaviour of block sliding by elegantly choosing simulation parameters; for example, strong (weak) interfaces have higher (lower) bond stiffness and, thus, can endure large (small) deformation before rupture, resulting in slow (fast) pulse-like fronts in the $\Tm$ case.

\subsection{Front occurrence independent of friction law}
As mentioned in the introduction, for explaining most of the frictional attributes in stiff solids such as metals, rocks, and glassy polymers, classical Coulomb and rate-and-state friction laws\cite{dieterich1979modeling,rice1983stability,scholz1998earthquakes} offer a satisfactory phenomenological explanation. However, on a small scale, its fundamental basis to represent the interface dynamics of macro contacts remains an active area for research\cite{baumberger2006solid, li2011frictional}.

On the contrary, the question of interface rupture mode for the contacting bodies transitioning from static to sliding is relevant for the extended interfaces and sliding of elastically complaint bodies. In such situations, the validity of friction laws is questionable\cite{schallamach1963theory, grosch1963relation, viswanathan2016stick}. Moreover, the observation of the moving fronts across various material systems (glasses\cite{rubinstein2004detachment}, rocks\cite{kaproth2013slow}, hydrogels\cite{baumberger2002self}) suggests that they likely occur for a wide variety of friction laws on a small scale such as slip-weakening\cite{ampuero2002nucleation}, and velocity-weakening\cite{brener2002frictional, gabriel2012transition, zheng1998conditions}. Our results are independent of any friction law, rather derived from fundamental macrocontact physics of detachment and reattachment governed by simple distance-based criterion.

\subsection{Existing theories for rupture mode selection}  
We now discuss various theories that have been proposed for pulse-like rupture modes. First, the most widely recognised theoretical explanation for self-healing cracks is the necessity of the velocity-weakening friction\cite{zheng1998conditions, heaton1990evidence}, which states that the friction stress at the contact interface decreases with the slip velocity $\dot{u}$. The crack heals when the slip velocity $\dot{u}$ tends to zero, and the static friction resumes. Alternative hypotheses include the need for stress barrier that reflects waves leading to crack healing\cite{johnson1990initiation} and normal load coupling at a bi-material interface with one body being stiffer than the other\cite{andrews1997wrinkle, cochard2000fault}.

In our numerical simulation, the pulse forms as a consequence of the macro-contacts physics, wherein both front speed and re-adhesion process determine whether the nucleated rupture will manifest as a stable pulse or an unstable crack. Apart from the intrinsic material properties (e.g., stiffness, breaking strength) of the interface, the front speeds, as described in the results, are also governed by the interface stress fields developed by the remote loading conditions. In our present analysis, re-adhesion occurs instantaneously once the prescribed distance criterion between opposing adhesive joints is satisfied. Thus, re-adhesion is rate-independent; otherwise, it is generally dynamic. In addition to the proximity requirement, if the slip velocity is also small enough for the re-adhesion process, crack healing becomes rate-dependent and thus influences rupture mode selection.

We note that there is no stress barrier along the interface in our simulations. Nevertheless, normal stress $\sigma_{yy}$ variation does indeed occur across a propagating pulse, likely due to the coupling between macrocontacts and the elasticity, resulting in local interface deformation. This normal coupling is not only attributable to the bi-material interface but also to the asymmetry in loading and sizes of the contacting bodies at the interface.

Recently, it has been shown that the pulse mode is favorable in displacement-controlled loading over an extended interface\cite{thogersen2021minimal}; however, for a finite interface, we demonstrated that a crack-like rupture might also be observed depending on how remote displacement load is applied.

\subsection{Analogy with the classical crack of fracture mechanics}
We have demonstrated that the transition from static to dynamic can occur via three different fronts in the same system depending on the remote loading conditions. Examining the accompanying stresses along the interface reveals that shear stress near the pulse-like front tip in both $\Tm$ and $\Tp$ cases resemble mode II shear crack (see also appendix \ref{sec:app_singular}). This analogy for a crack-like front in $\To$ case cannot be accurately quantified as the shear stress exhibits very sharp peak near the leading rupture tip (see Fig. \ref{fig:top_stress}(b)). Nevertheless, it is shown in a qualitative sense that this crack-like front also experiences mode I loading, see Fig. \ref{fig:top_stress}(c), making a strong case for analogies with mixed mode interface fracture. 

Our simulation also manages to capture the transition in front speed: from slow to fast in $\Tm$ and $\Tp$ (Fig.~\ref{fig:left_fronts}(c) and \ref{fig:right_fronts}(c)) and a rapid acceleration to Rayleigh speed in the $\To$ case (Fig. \ref{fig:top_fronts}(c)). For our chosen parameters, the maximum front speed is observed to be Rayleigh wave speed, consistent with the maximum speed that a crack can attain in homogeneous material\cite{freund1998dynamic}. In scenarios with a weakened interface, a fast-moving front might trigger a secondary front ahead of it and potentially transitioning to supershear speed -- a well-recognized Burridge-Andrews mechanism\cite{andrews1976rupture, xia2004laboratory} behind supershear front speed can also be investigated under this framework.

Note that the simulation results for a larger lattice, maintaining the same aspect ratio but $l=300$ contacts point, retains similar qualitative behaviour for all loading conditions discussed. However, for $l = 500$ contact points, as mentioned for the $\To$ case, a front rapidly breaks all interface bonds. The reformation process of large contact nodes occurs unevenly, with some regions of interface bonds taking considerable time to rebond, as we saw in Fig.~\ref{fig:top_fronts}(b), thus resulting in an inhomogeneous interface. This new inhomogeneous interface can alter the direction and local rupture mode. But, subsequently, the transition from static to dynamic motion occurs via crack-like rupture mode.

\subsection{Analogy with locomotory waves in soft invertebrates}
Another curious analogy can be drawn between the three rupture modes in our model and locomotory wave in soft-bodied organisms. These individual possess elastic bodies that effect coordinated motion of different body segments along the moving direction. For locomotion, these organisms have specialized muscles that provide compliance to aid in locomotory wave propagation. Direct contraction waves, such as those observed in gastropods (e.g., \emph{M. labio} form \emph{confusa}\cite{kuroda2014common}) locomotion, propagate directly from the posterior to the anterior during forward motion. This resembles the pulses in $\Tm$ case (\emph{c.f. \ref{sec:push}}).

Similar to the pulse moving against the sliding direction as observed in $\Tp$ case, the chitons (e.g., \emph{I. comptus}\cite{kuroda2014common}) utilize retrograde waves for forward motion. The anterior edge of the chitons extends in an anterior direction that travels towards the posterior end. The retrograde locomotory waves are also seen in some Nemertines, such as \emph{Rhynchodemus}\cite{pantin1950locomotion}, where the body segments lose contact with the ground when relaxed and establish contact points in the regions of contraction. The organism forms a tensile neck at its head, and the extended muscles in this region induce interfacial separation from the ground. This neck is then propagated posteriorly as a retrograde wave.

\section{Conclusions}
This work demonstrated that the transition from static to dynamic sliding occurs via three different rupture fronts in the same adhesive frictional interface between finite-size deformable bodies,  depending on how the remote boundary loading is applied. We specifically discuss three types of velocity-controlled loading---push ($\Tm$), pull ($\Tp$), and slide ($\To$). To explore these loading conditions, a numerical lattice-spring network is formulated to obtain the two-dimensional interface deformation and contacting bodies.

In $\Tm$ and $\To$ cases, the onset of front initiation is determined by the shear and normal stresses, while in $\Tp$ case, it is mainly influenced by the shear stress. Once the front initiates, its mode selection depends on how the stress distributes across the interface. The shear stress is confined near the front tip region in both $\Tm, \Tp$ cases, while far away, the interface remains unstressed---this results in slow fronts as it requires time for the interface far from the tip to bear the applied load. This time obviously depends on sliding speed $v_0$ as we saw that the front speeds scale linearly with it. These slow fronts are likely to manifest as pulses because of the sufficient time available for the bonds to reform at the front trailing end. Moreover, the speed of these pulses is also affected by the normal stress, as we have seen for the $\Tp$ case. 

In contrast, the entire interface accumulates finite shear stress in the $\To$ case. The rupture front gains speed by relieving this accumulated stress, resulting in a fast crack-like front.

\counterwithin{figure}{section}
\appendix
\section*{\centering\Large Appendices} 
\addcontentsline{toc}{section}{Appendices}

\section{Variation of pulse width}
\begin{figure}[ht!]
	\centering	
	\includegraphics[width=0.5\linewidth]{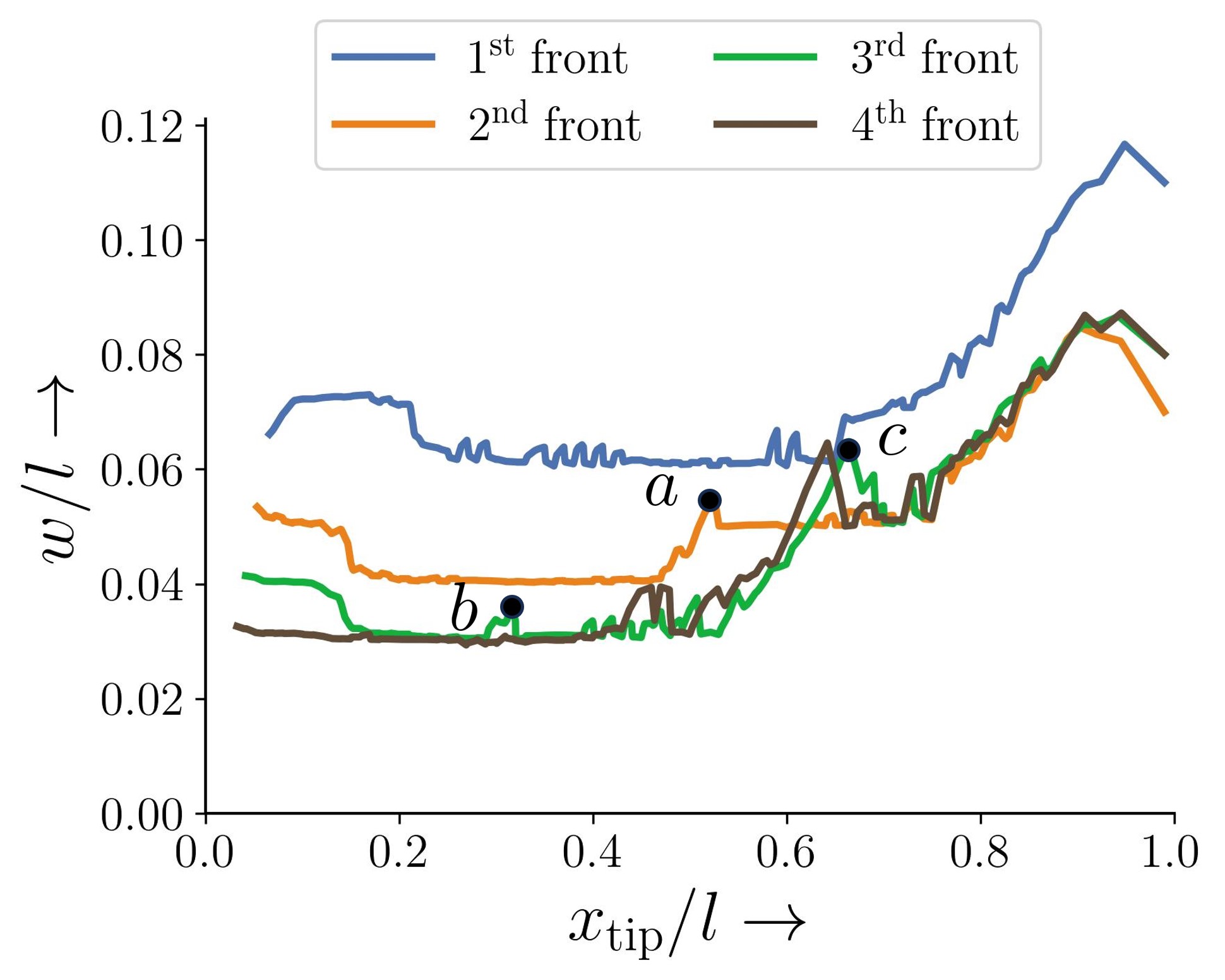}
	\caption{Pulse width variation for pushing case. The driving velocity $v_0$, normal load $F_N$ and bond parameters are same as manuscript Fig. \ref{fig:left_fronts}.}		
	\label{fig:app_pw}
\end{figure}

The plot in Fig. \ref{fig:app_pw} shows the width variation of the initial four sequential pulses, each represented by a distinct color, across the entire interface. During the initial propagation phase, the pulse width remains almost constant within a one-unit spacing (i.e., $w/l = 0.01$).
As the pulse nears the opposite boundary, its width tends to expand. Instances of abrupt changes in the pulse width (such as instances $a$, $b$, and $c$ labelled in the plot) occur due to the acceleration of the leading pulse near the opposite boundary, as explained in detail in the main text, section \ref{sec:push}. This acceleration causes a subsequent increase in the trailing pulses' widths. While the spatiotemporal plot of intact bonds in the main Fig. \ref{fig:right_fronts}(b) suggests a similar trend in pulse width for pulling, this information has been omitted for conciseness.

\section{Asymptotic singular shear stress during pulse motion}\label{sec:app_singular}
In pushing, the localised shear stress $\sigma_{xy}$ in the immediate vicinity of the pulse tip conforms to the asymptotic square-root singular solution of the fracture mechanics.
\begin{figure}[ht!]
	\centering	
	\includegraphics[width=\linewidth]{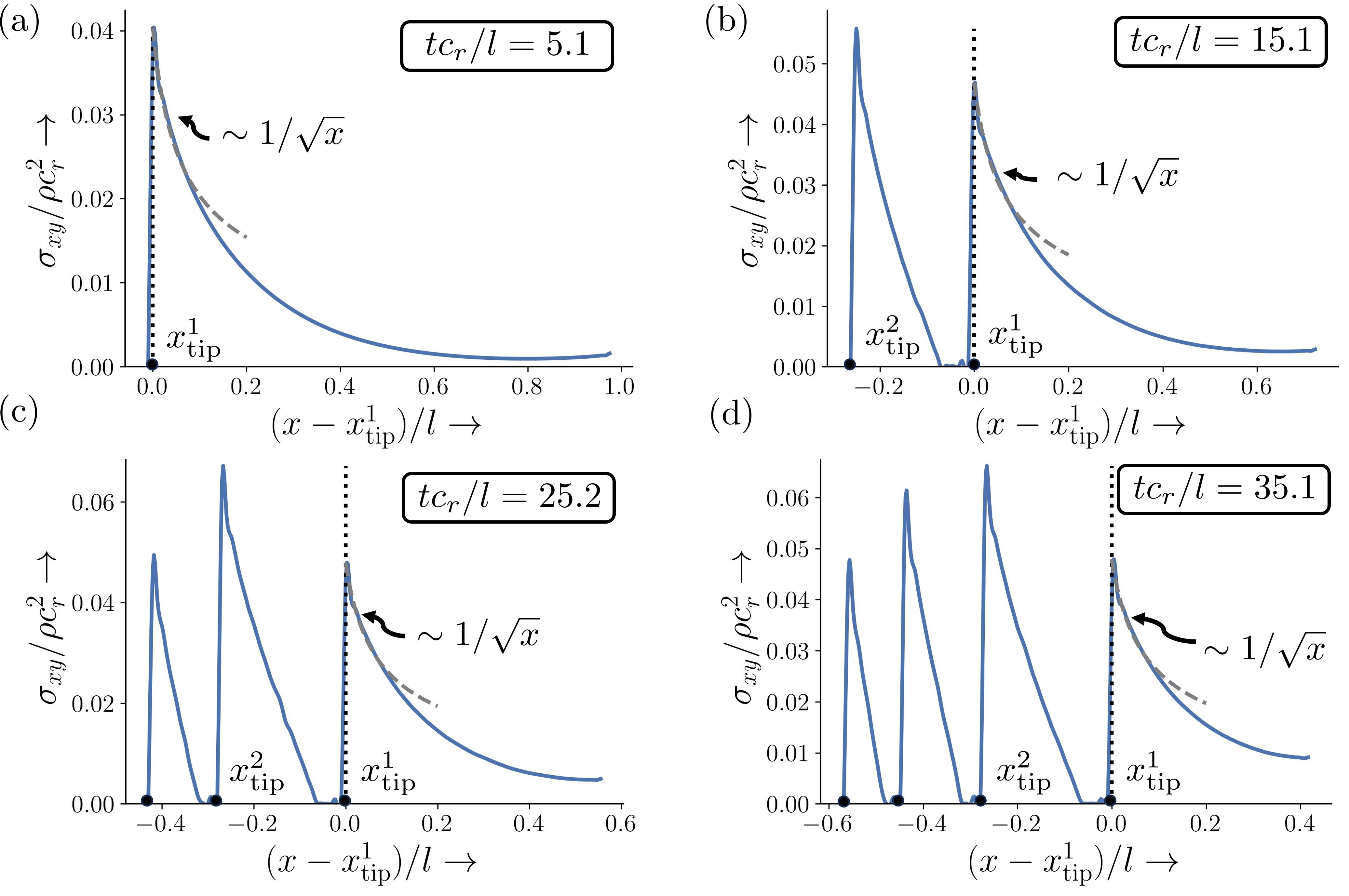}
	\caption{Fitting of an analytical singularity solution near the crack tip for first front of Fig. \ref{fig:left_stress}(b). The driving velocity $v_0$, normal load $F_N$ and bond parameters are same as manuscript Fig. \ref{fig:left_fronts}.}		
	\label{fig:sq_root1}
\end{figure} Fig. \ref{fig:sq_root1}(a)-(d) overplots the analytical singular solution (dashed curve) at the tip onto the numerical shear stress (blue curve) of the first pulse at four different time instances. Since the numerical solution is finite at the crack tip, the analytical solution is displaced by some distance $x_p$ given by
\begin{align}
	\sigma_{xy} = \frac{K_{II}}{\sqrt{2\pi (x+x_p)}}
\end{align} 
that involves two parameters $K_{II}$ and $x_p$, which can be correlated to simplify into a single fitting parameter by matching the maximum finite stress value at $x=0$ as 
\begin{align}
	\sigma_{xy} = \frac{K_{II}}{\sqrt{2\pi x_p}} = \sigma_{xy}^{\text{max}}
\end{align}

\begin{figure}[ht!]
	\centering	
	\includegraphics[width=\linewidth]{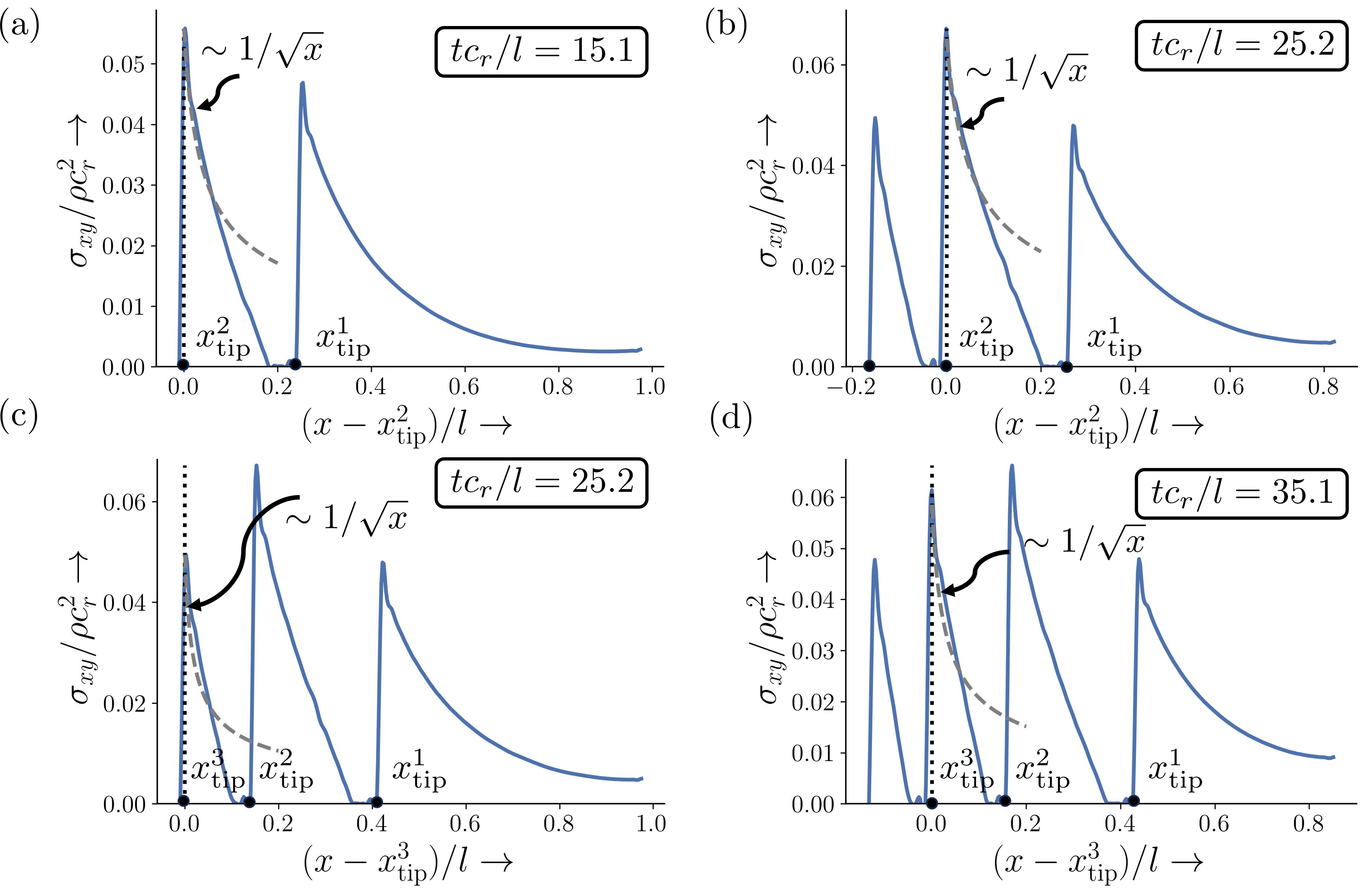}
	\caption{Fitting of an analytical solution near the pulse tip for second and third pulse of Fig. \ref{fig:left_stress}(b). The driving velocity $v_0$, normal load $F_N$ and bond parameters are same as manuscript Fig. \ref{fig:left_fronts}.}		
	\label{fig:sq_root23}
\end{figure}

In Fig. \ref{fig:sq_root23}(a) and (b), a square-root fitting is demonstrated for the second pulse (in (a)-(b)) and the third pulse (in (c)-(d)). It's reasonable to expect similar outcomes in the case of pulling.

\section{Horizontal slip velocity $\dot{u}$ at three locations along the interface}\label{sec:app_slipvel}
\begin{figure}[ht!]
	\centering	
	\includegraphics[width=0.6\linewidth]{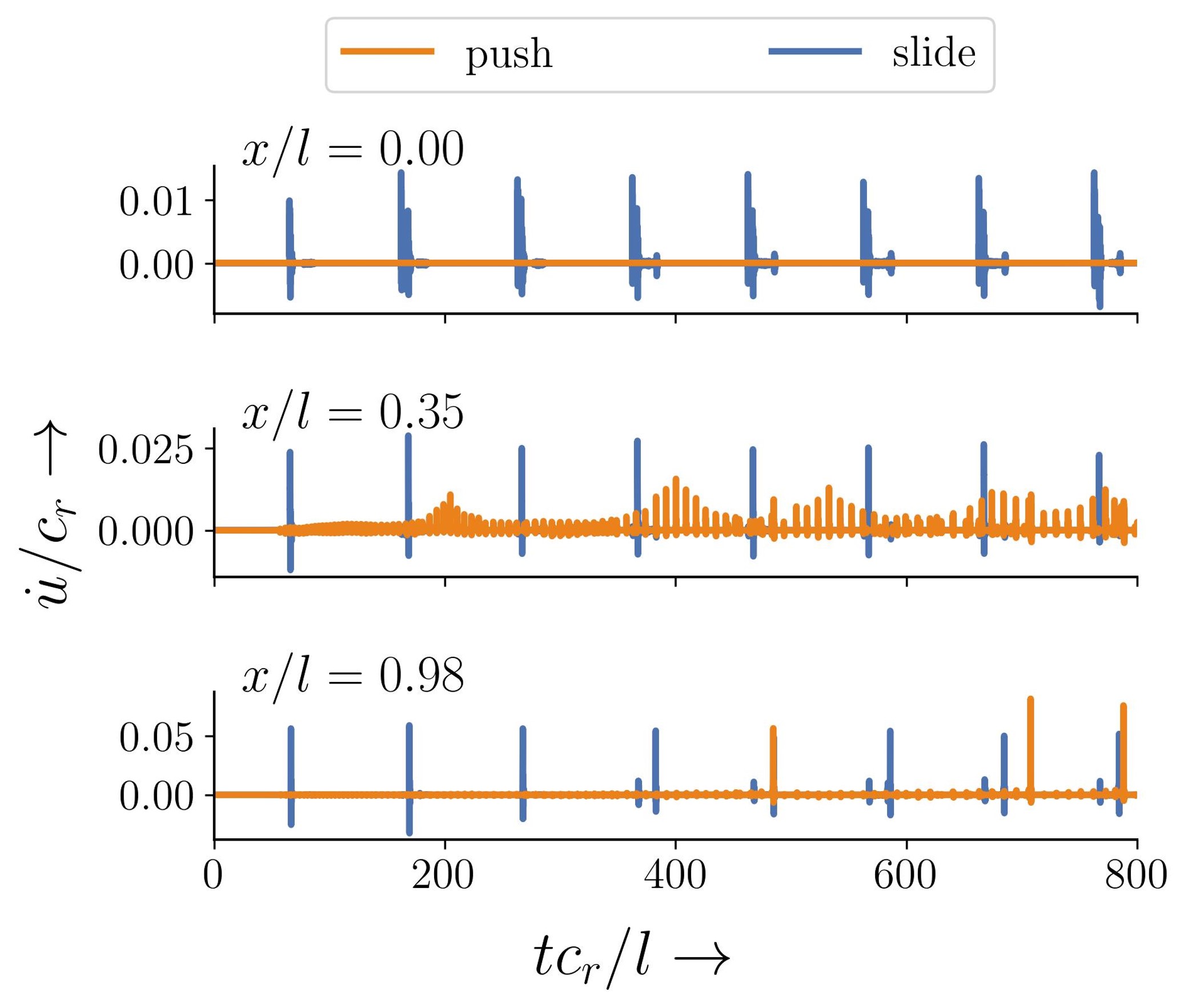}
	\caption{(a) Horizontal slip velocity ($\dot{u}$) of three different interface locations ($x/l = 0, 0.35,$ and $0.98$) for rear-pushing (orange curves) and top (blue curves) sliding configurations. The driving velocity $v_0$, normal load $F_N$ and bond parameters are same as manuscript Fig. \ref{fig:uslip}.}		
	\label{fig:app_uslipvel}
\end{figure}

As mentioned in the main text in the section \ref{sec:slip}, an additional differentiation between the body movements under various modes can be observed by examining the slip velocity $\dot{u}$. Fig. \ref{fig:app_uslipvel} compares the horizontal slip velocity $\dot{u}$ at three distinct points along the interface for pushing and sliding scenarios. The distinction between the pulse-like and crack-like motion is evident from the figure. In the case of pushing, a gradual variation in slip velocity $\dot{u}$ characterizes the formation of a slow slip pulse (see the orange curve in a middle panel in Fig. \ref{fig:app_uslipvel}). At any midpoint (say $x/l=0.35$), $\dot{u}$ increases upon the pulse's arrival and decreases upon its departure. The point situated at the opposite edge ($x/l=0.98$) remains stationary until the advancing pulse, which becomes sufficiently fast when it arrives at that location (refer to Fig. \ref{fig:left_fronts}(c)), induces a sudden jump in slip velocity. 
  
In the other case of sliding, the developed crack-like front being fast induces a sharp peak in slip velocity at all three locations, as seen blue curves in Fig. \ref{fig:app_uslipvel}.

\section{Pulses speed for pushing versus pulling scenario}\label{sec:app_speed}
\begin{figure}[ht!]
	\centering	
	\includegraphics[width=\linewidth]{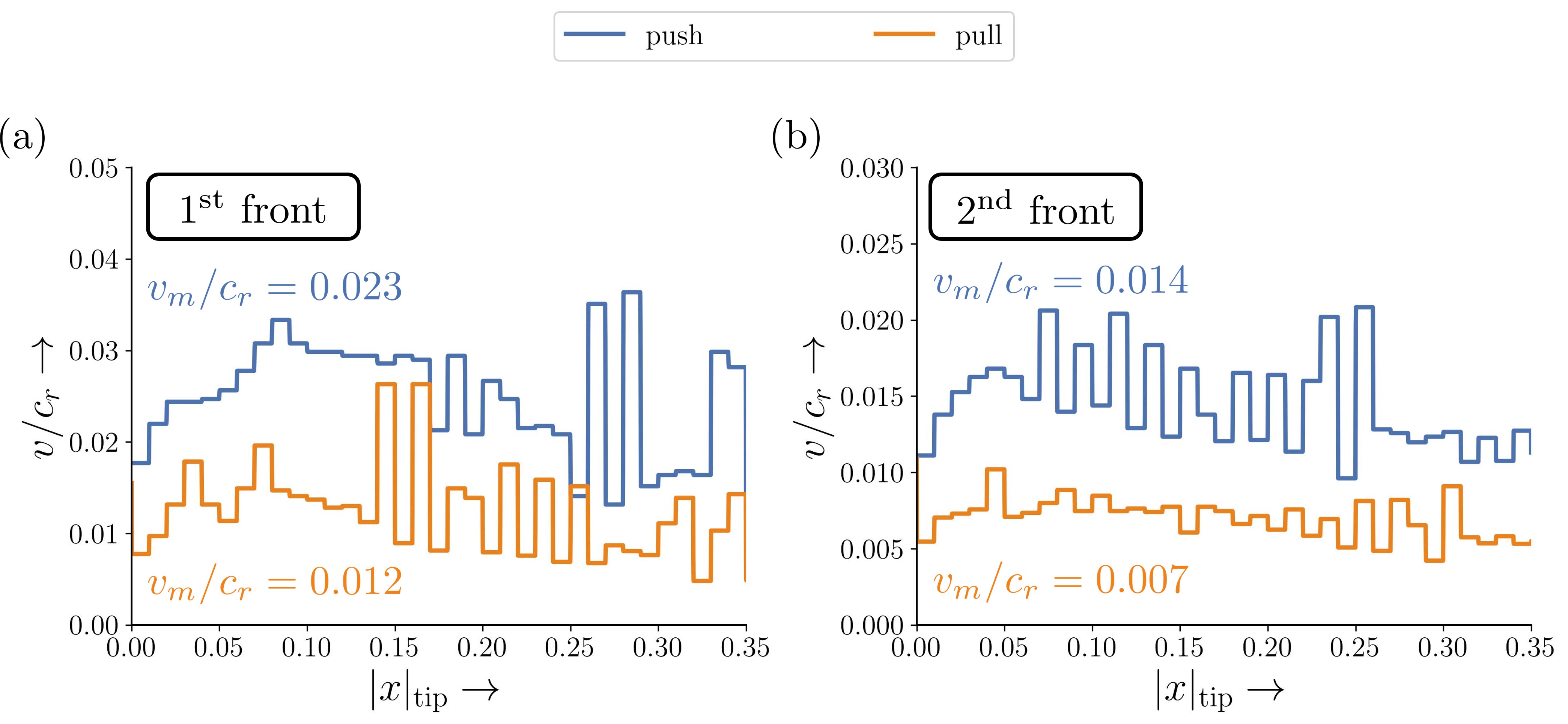}
	\caption{Comparison of the speed of the (a) first and the (b) second pulse fronts in two different sliding scenarios. The driving velocity $v_0$, normal load $F_N$ and bond parameters are same as manuscript Fig. \ref{fig:left_fronts}.}		
	\label{fig:app_speed_lr}
\end{figure}
We compare the speed of the pulses for two loading scenarios - pushing and pulling, both with a constant sliding velocity $v_0$. As discussed in the manuscript section, the pushing produces pulses in the direction of sliding, while in pulling, the pulses move in the opposite direction to the gross sliding. Fig. \ref{fig:app_speed_lr} compares the speed of the first two pulses. It can be identified that the pulses are $50\%$ slower in pulling as compared to pushing. The distinction arises due to the more prominently compressive nature of $\sigma_{yy}$ at nucleation in pulling compared to pushing. Additionally, the mean speeds are annotated by the respective colour in the plot.

\clearpage
\bibliography{bibfile}

\begin{thebibliography}{10}

\bibitem{rabinowicz1956stick}
Rabinowicz E.
\newblock Stick and slip.
\newblock Scientific American. 1956;194(5):109--119.

\bibitem{rabinowicz1958intrinsic}
Rabinowicz E.
\newblock The intrinsic variables affecting the stick-slip process.
\newblock Proceedings of the Physical Society. 1958;71(4):668.

\bibitem{brace1966stick}
Brace W, Byerlee J.
\newblock Stick-slip as a mechanism for earthquakes.
\newblock Science. 1966;153(3739):990--992.

\bibitem{scholz1998earthquakes}
Scholz CH.
\newblock Earthquakes and friction laws.
\newblock Nature. 1998;391(6662):37--42.

\bibitem{bowden2001friction}
Bowden FP, Tabor D.
\newblock The friction and lubrication of solids. vol.~1.
\newblock Oxford university press; 2001.

\bibitem{dieterich1979modeling}
Dieterich JH.
\newblock Modeling of rock friction: 1. Experimental results and constitutive
  equations.
\newblock Journal of Geophysical Research: Solid Earth. 1979;84(B5):2161--2168.

\bibitem{dieterich1978time}
Dieterich JH.
\newblock Time-dependent friction and the mechanics of stick-slip.
\newblock Rock friction and earthquake prediction. 1978;p. 790--806.

\bibitem{ruina1983slip}
Ruina A.
\newblock Slip instability and state variable friction laws.
\newblock Journal of Geophysical Research: Solid Earth.
  1983;88(B12):10359--10370.

\bibitem{achenbach1967dynamic}
Achenbach JD, Epstein HI.
\newblock Dynamic interaction of a layer and a half-space.
\newblock Journal of the Engineering Mechanics Division. 1967;93(5):27--42.

\bibitem{rubinstein2004detachment}
Rubinstein SM, Cohen G, Fineberg J.
\newblock Detachment fronts and the onset of dynamic friction.
\newblock Nature. 2004;430(7003):1005--1009.

\bibitem{gvirtzman2021nucleation}
Gvirtzman S, Fineberg J.
\newblock Nucleation fronts ignite the interface rupture that initiates
  frictional motion.
\newblock Nature Physics. 2021;17(9):1037--1042.

\bibitem{xia2004laboratory}
Xia K, Rosakis AJ, Kanamori H.
\newblock Laboratory earthquakes: The sub-Rayleigh-to-supershear rupture
  transition.
\newblock Science. 2004;303(5665):1859--1861.

\bibitem{lu2007pulse}
Lu X, Lapusta N, Rosakis AJ.
\newblock Pulse-like and crack-like ruptures in experiments mimicking crustal
  earthquakes.
\newblock Proceedings of the National Academy of Sciences.
  2007;104(48):18931--18936.

\bibitem{viswanathan2016stick}
Viswanathan K, Sundaram NK, Chandrasekar S.
\newblock Stick-slip at soft adhesive interfaces mediated by slow frictional
  waves.
\newblock Soft matter. 2016;12(24):5265--5275.

\bibitem{ben2010dynamics}
Ben-David O, Cohen G, Fineberg J.
\newblock The dynamics of the onset of frictional slip.
\newblock Science. 2010;330(6001):211--214.

\bibitem{shlomai2016structure}
Shlomai H, Fineberg J.
\newblock The structure of slip-pulses and supershear ruptures driving slip in
  bimaterial friction.
\newblock Nature communications. 2016;7(1):11787.

\bibitem{mindlin1949compliance}
Mindlin RD.
\newblock Compliance of elastic bodies in contact.
\newblock Journal of Applied Mechanics. 1949;.

\bibitem{johnson1971surface}
Johnson KL, Kendall K, Roberts a.
\newblock Surface energy and the contact of elastic solids.
\newblock Proceedings of the royal society of London A mathematical and
  physical sciences. 1971;324(1558):301--313.

\bibitem{savkoor1977effect}
Savkoor A, Briggs G.
\newblock The effect of tangential force on the contact of elastic solids in
  adhesion.
\newblock Proceedings of the Royal Society of London A Mathematical and
  Physical Sciences. 1977;356(1684):103--114.

\bibitem{adams2014stick}
Adams GG.
\newblock Stick, partial slip and sliding in the plane strain micro contact of
  two elastic bodies.
\newblock Royal Society open science. 2014;1(3):140363.

\bibitem{heaton1990evidence}
Heaton TH.
\newblock Evidence for and implications of self-healing pulses of slip in
  earthquake rupture.
\newblock Physics of the Earth and Planetary Interiors. 1990;64(1):1--20.

\bibitem{zheng1998conditions}
Zheng G, Rice JR.
\newblock Conditions under which velocity-weakening friction allows a
  self-healing versus a cracklike mode of rupture.
\newblock Bulletin of the Seismological Society of America.
  1998;88(6):1466--1483.

\bibitem{johnson1990initiation}
Johnson E.
\newblock On the initiation of unidirectional slip.
\newblock Geophysical Journal International. 1990;101(1):125--132.

\bibitem{andrews1997wrinkle}
Andrews DJ, Ben-Zion Y.
\newblock Wrinkle-like slip pulse on a fault between different materials.
\newblock Journal of Geophysical Research: Solid Earth. 1997;102(B1):553--571.

\bibitem{cochard2000fault}
Cochard A, Rice J.
\newblock Fault rupture between dissimilar materials: Ill-posedness,
  regularization, and slip-pulse response.
\newblock Journal of Geophysical Research: Solid Earth.
  2000;105(B11):25891--25907.

\bibitem{tromborg2011transition}
Tr{\o}mborg J, Scheibert J, Amundsen DS, Th{\o}gersen K, Malthe-S{\o}renssen A.
\newblock Transition from static to kinetic friction: insights from a 2D model.
\newblock Physical review letters. 2011;107(7):074301.

\bibitem{thogersen2021minimal}
Th{\o}gersen K, Aharonov E, Barras F, Renard F.
\newblock Minimal model for the onset of slip pulses in frictional rupture.
\newblock Physical Review E. 2021;103(5):052802.

\bibitem{gerde2001friction}
Gerde E, Marder M.
\newblock Friction and fracture.
\newblock Nature. 2001;413(6853):285--288.

\bibitem{braun2009dynamics}
Braun O, Barel I, Urbakh M.
\newblock Dynamics of transition from static to kinetic friction.
\newblock Physical review letters. 2009;103(19):194301.

\bibitem{ansari2022propagating}
Ansari MA, Viswanathan K.
\newblock Propagating Schallamach-type waves resemble interface cracks.
\newblock Physical Review E. 2022;105(4):045002.

\bibitem{burridge1967model}
Burridge R, Knopoff L.
\newblock Model and theoretical seismicity.
\newblock Bulletin of the seismological society of america.
  1967;57(3):341--371.

\bibitem{amundsen20121d}
Amundsen DS, Scheibert J, Th{\o}gersen K, Tr{\o}mborg J, Malthe-S{\o}renssen A.
\newblock 1D model of precursors to frictional stick-slip motion allowing for
  robust comparison with experiments.
\newblock Tribology Letters. 2012;45:357--369.

\bibitem{schallamach1971does}
Schallamach A.
\newblock How does rubber slide?
\newblock Wear. 1971;17(4):301--312.

\bibitem{viswanathan2022fifty}
Viswanathan K, Chandrasekar S.
\newblock Fifty years of Schallamach waves: From rubber friction to nanoscale
  fracture.
\newblock Philosophical Transactions of the Royal Society A.
  2022;380(2232):20210339.

\bibitem{svetlizky2017brittle}
Svetlizky I, Kammer DS, Bayart E, Cohen G, Fineberg J.
\newblock Brittle fracture theory predicts the equation of motion of frictional
  rupture fronts.
\newblock Physical review letters. 2017;118(12):125501.

\bibitem{martin2000dynamic}
Mart{\'\i}n T, Espanol P, Rubio MA, Z{\'u}niga I.
\newblock Dynamic fracture in a discrete model of a brittle elastic solid.
\newblock Physical review E. 2000;61(6):6120.

\bibitem{del2002wave}
Del Valle-Garcia R, Sanchez-Sesma F.
\newblock Wave scattering effects in elastic percolation models.
\newblock Molecular Physics. 2002;100(19):3167--3172.

\bibitem{braun2014new}
Braun M, Fern{\'a}ndez-S{\'a}ez J.
\newblock A new 2D discrete model applied to dynamic crack propagation in
  brittle materials.
\newblock International Journal of Solids and Structures.
  2014;51(21-22):3787--3797.

\bibitem{schallamach1963theory}
Schallamach A.
\newblock A theory of dynamic rubber friction.
\newblock Wear. 1963;6(5):375--382.

\bibitem{filippov2004friction}
Filippov A, Klafter J, Urbakh M.
\newblock Friction through dynamical formation and rupture of molecular bonds.
\newblock Physical Review Letters. 2004;92(13):135503.

\bibitem{rice1983stability}
Rice J.
\newblock Stability of Steady Frictional Slipping.
\newblock Journal of Applied Mechanics. 1983;50(2):343--349.

\bibitem{baumberger2006solid}
Baumberger T, Caroli C.
\newblock Solid friction from stick--slip down to pinning and aging.
\newblock Advances in Physics. 2006;55(3-4):279--348.

\bibitem{li2011frictional}
Li Q, Tullis TE, Goldsby D, Carpick RW.
\newblock Frictional ageing from interfacial bonding and the origins of rate
  and state friction.
\newblock Nature. 2011;480(7376):233--236.

\bibitem{grosch1963relation}
Grosch K.
\newblock The relation between the friction and visco-elastic properties of
  rubber.
\newblock Proceedings of the Royal Society of London Series A Mathematical and
  Physical Sciences. 1963;274(1356):21--39.

\bibitem{kaproth2013slow}
Kaproth BM, Marone C.
\newblock Slow earthquakes, preseismic velocity changes, and the origin of slow
  frictional stick-slip.
\newblock Science. 2013;341(6151):1229--1232.

\bibitem{baumberger2002self}
Baumberger T, Caroli C, Ronsin O.
\newblock Self-healing slip pulses along a gel/glass interface.
\newblock Physical review letters. 2002;88(7):075509.

\bibitem{ampuero2002nucleation}
Ampuero JP, Vilotte JP, Sanchez-Sesma F.
\newblock Nucleation of rupture under slip dependent friction law: simple
  models of fault zone.
\newblock Journal of Geophysical Research: Solid Earth. 2002;107(B12):ESE--2.

\bibitem{brener2002frictional}
Brener EA, Marchenko VI.
\newblock Frictional shear cracks.
\newblock Journal of Experimental and Theoretical Physics Letters.
  2002;76:211--214.

\bibitem{gabriel2012transition}
Gabriel AA, Ampuero JP, Dalguer L, Mai PM.
\newblock The transition of dynamic rupture styles in elastic media under
  velocity-weakening friction.
\newblock Journal of Geophysical Research: Solid Earth. 2012;117(B9).

\bibitem{freund1998dynamic}
Freund LB.
\newblock Dynamic fracture mechanics.
\newblock Cambridge university press; 1998.

\bibitem{andrews1976rupture}
Andrews D.
\newblock Rupture velocity of plane strain shear cracks.
\newblock Journal of Geophysical Research. 1976;81(32):5679--5687.

\bibitem{kuroda2014common}
Kuroda S, Kunita I, Tanaka Y, Ishiguro A, Kobayashi R, Nakagaki T.
\newblock Common mechanics of mode switching in locomotion of limbless and
  legged animals.
\newblock Journal of the Royal Society interface. 2014;11(95):20140205.

\bibitem{pantin1950locomotion}
Pantin C.
\newblock Locomotion in British terrestrial nemertines and planarians: with a
  discussion on the identity of Rhynchodemus bilineatus (Mecznikow) in Britain,
  and on the name Fasciola terrestris OF M{\"u}ller.
\newblock In: Proceedings of the Linnean Society of London. vol. 162. Wiley
  Online Library; 1950. p. 23--37.

\end{thebibliography}
\bibliographystyle{vancouver}

\end{document}